\documentclass[fleqn,10pt]{wlscirep}

\usepackage{braket}
\usepackage{amsmath}
\usepackage{float}

\title{Ordered states in the Kitaev-Heisenberg model: From 1D chains to 2D honeycomb}

\author[1,*]{Cli\`o Efthimia Agrapidis}
\author[1,2]{Jeroen van den Brink}
\author[1,2]{Satoshi Nishimoto}
\affil[1]{IFW Dresden, Institute for Theoretical Solid State Physics, Dresden, 01069, Germany}
\affil[2]{Technical University, Department of Physics, Dresden, 01069, Germany}

\affil[*]{c.agrapidis@ifw-dresden.de}

\begin{abstract}
We study the ground state of the 1D Kitaev-Heisenberg (KH) model using the density-matrix renormalization group and Lanczos exact diagonalization methods. We obtain a rich ground-state phase diagram as a function of the ratio between Heisenberg ($J=\cos\phi)$ and Kitaev ($K=\sin\phi$) interactions. Depending on the ratio, the system exhibits four long-range ordered states: ferromagnetic-$z$ , ferromagnetic-$xy$, staggered-$xy$, N\'eel-$z$, and two liquid states: Tomonaga-Luttinger liquid and spiral-$xy$. The two Kitaev points $\phi=\frac{\pi}{2}$ and $\phi=\frac{3\pi}{2}$ are singular. The $\phi$-dependent phase diagram is similar to that for the 2D honeycomb-lattice KH model. Remarkably, all the ordered states of the honeycomb-lattice KH model can be interpreted in terms of the coupled KH chains. We also discuss the magnetic structure of the K-intercalated RuCl$_3$, a potential Kitaev material, in the framework of the 1D KH model. Furthermore, we demonstrate that the low-lying excitations of the 1D KH Hamiltonian can be explained within the combination of the known six-vertex model and spin-wave theory.
\end{abstract}
\begin{document}

\flushbottom
\maketitle
\thispagestyle{empty}

%\noindent Please note: Abbreviations should be introduced at the first mention in the main text no abbreviations lists. Suggested structure of main text (not enforced) is provided below.

\section*{Introduction}

It is well-known that magnetic ordering (e.g. the N\'eel state) is induced by the spontaneous breaking of some sort of symmetry at low temperatures. If one can prevent such a {\it conventional} magnetic ordering, an exotic type of quantum-disordered state --- the so-called {\it quantum spin liquid} (QSL) --- may arise. In the spin liquid phase, a macroscopic number of quasi-degenerate low-energy states compete with each other. Since Anderson's proposal of a resonating valence bond state as a theoretical realization of a QSL in 1973~\cite{Anderson73}, geometrical frustration~\cite{Moessner06}, a situation where the network of magnetic interactions is incompatible with the spatial symmetry, had been considered to be almost the key to achieve such situation~\cite{Balents10}. But in 2006, the advent of the Kitaev model~\cite{Kitaev06} shifted the focus on another way to realize a QSL --- frustration due to bond-dependent exchange interactions~\cite{Nussinov15}. This model has nontrivial topological phases with elementary excitations exhibiting a Majorana algebra. After that, the microscopic origin of such Kitaev-type interactions in the $d^5$ transition metal compounds was worked out~\cite{Khaliullin05,Jackeli09}. In the last decade, there has been a growing number of researches on the Kitaev materials~\cite{Trebst17, Savary16}.

The first Kitaev material candidates are 5$d^5$ honeycomb iridates $A_2$IrO$_3$ ($A$=Na or Li)~\cite{Jackeli09,Chaloupka10}. The Ir$^{4+}$ ions centered in the edge-sharing IrO$_6$ octahedra form a planar hexagonal network. In each Ir$^{4+}$ ion, the two $e_{\rm g}$ levels are split off by the crystal field and five electrons in the 5$d$ shell are put into the $t_{\rm 2g}$ orbitals with a finite effective angular momentum $l=1$. Besides, the strong spin-orbit coupling (SOC) for 5$d$ electrons splits up the $t^5_{\rm 2g}$ manifold into a fully occupied $j=\frac{3}{2}$ quartet band and a half-filled $j=\frac{1}{2}$ doublet band. The latter half-filled band gives an effective $j=\frac{1}{2}$ pseudospin model, in association with the opening of a Mott gap. In the edge-sharing IrO$_6$ octahedra, there are two Ir-O-Ir exchange paths with a $\sim90^\circ$ bond angle. The anisotropic Kitaev-type coupling then arises from the particular superexchange interactions between the $j=\frac{1}{2}$ pseudospins. A more realistic spin model to describe the magnetic properties is the Kitaev-Heisenberg (KH) model, that accounts for the residual Heisenberg-type couplings. It is known that the Kitaev QSL is most likely preempted by a certain level of the Heisenberg interaction. In fact, the above iridates exhibit long-range magnetic order at low temperatures~\cite{Singh10,Ye12,Choi12,Takayama15}, possibly assisted by off-diagonal and/or long-range interactions. 

At present, one of the most promising materials close to the Kitaev QSL is the 4$d^5$ ruthenium trichloride $\alpha$-RuCl$_3$ in its honeycomb crystal phase~\cite{Plumb14,Rousochatzakis15,Kim15,Sears15,Majumder15,Kubota15,Nasu16}. For this compound the effective $j=\frac{1}{2}$ description, as a Ru$^{3+}$ 4$d$ analogue to the iridates, seems to still give a good description of the magnetic properties, although some degree of admixture between $j=\frac{1}{2}$ and $j=\frac{3}{2}$ in the $t^5_{\rm 2g}$ states may arise from a weaker SOC for 4$d$ electrons than in the Ir 5$d$ orbitals. Neutron and Raman scattering studies gave evidence for fractionalized excitations typical of the Kitaev QSL~\cite{Banerjee16,Sandilands15,Nasu17}, and both theoretical~\cite{Ravi16} and experimental investigations~\cite{Baek17,Anja17,Hentrich17,Zheng17,Leahy16,Zhou16} indicated, for this material, the existence of a transition into a QSL in the presence of external magnetic field. Furthermore, a recent X-ray diffraction measurement suggested the Kitaev QSL is induced by a small applied pressure of about 0.7 GPa~\cite{Wang17}. A possibility of pressure-induced QSL has been also reported for the other Kitaev materials, honeycomb iridates $\beta$-Li$_2$IrO$_3$~\cite{Veiga17} and $\gamma$-Li$_2$IrO$_3$~\cite{Breznay17}. In both materials the low temperature phase is magnetically ordered without field and under ambient pressure. Namely, the Kitaev QSL is derived through a melting of the magnetic ordering by field or pressure. Therefore, it is crucial to deeply understand the characteristics of the magnetically ordered states.

Very recently, a temperature-dependent electron energy loss spectroscopy (EELS) measurement was performed in a K-intercalated RuCl$_3$, denoted as K$_{0.5}$RuCl$_3$ (Knupfer M., private communications). The intercalated K$^+$ ions provide charge carriers, however, a sharp gap was observed at $\sim 0.9$ eV, instead of  the charge gap $E_{\rm g}=1.1-1.2$ eV for the undoped $\alpha$-RuCl$_3$~\cite{Koitzsch16,Sandilands16-1,Sandilands16-2}. This indicates an insulating feature of K$_{0.5}$RuCl$_3$ and differs from the pseudogap behavior seen for charge localization in disordered metals. This has been interpreted as half of the $j=\frac{1}{2}$ pseudospins being replaced by nonmagnetic $d^6$ ions. Therefore, this insulating state can be pictured as a formation of superlattice by charge disproportionation (charge ordering). Different possible charge ordering patterns are shown in Fig.~\ref{fig_lattice}. Of particular interest in the present context is the zigzag chain-type pattern exhibited in Fig.~\ref{fig_lattice}(c), where chains of nonmagnetic ions are separated by chains of magnetic ions with KH interactions.

The above motivations lead us to studying the 1D KH model not only for fundamental theoretical reasons, but also as a possible minimal spin model to describe the magnetic properties of the K-intercalated RuCl$_3$. So far, there are few studies on the 1D KH model~\cite{Sela14,Brzezicki07,Mondal08,Divakaran09,Eriksson09,Mahdavifara10,Subrahmanyam13,Katsura15,Steinigeweg16}. Using the density-matrix renormalization group (DMRG) method, we calculate total spin, real-space spin-spin correlation functions, static spin structure factor, central charge, and various order parameters in the ground state. Based on the results, we obtain the ground-state phase diagram, including four long-range ordered and two liquid phases, as a function of the ratio between Heisenberg and Kitaev interactions. Moreover, the relevance of the phase diagram to that of the honeycomb-lattice KH model is discussed. \textit{It is striking that all the magnetically ordered states of the honeycomb-lattice KH model can be interpreted in terms of coupled 1D KH chains.} Furthermore, we calculate the dynamical spin structure factors via the Lanczos exact diagonalization (ED) technique. The basic low-lying excitations are considered by use of the known six-vertex model and the spin-wave theory (SWT). The present investigation can thus contribute to an elucidation of the fundamental properties of the K-intercalated RuCl$_3$ as well as a better insight in the understanding of the physics of the honeycomb-lattice KH model.

\section*{Model and Numerical Methods}
\label{modelandmethod}

\subsection*{Pattern of charge ordering in the K-intercalated RuCl$_3$}

\begin{figure}[H]
\centering
\includegraphics[width=0.70\linewidth]{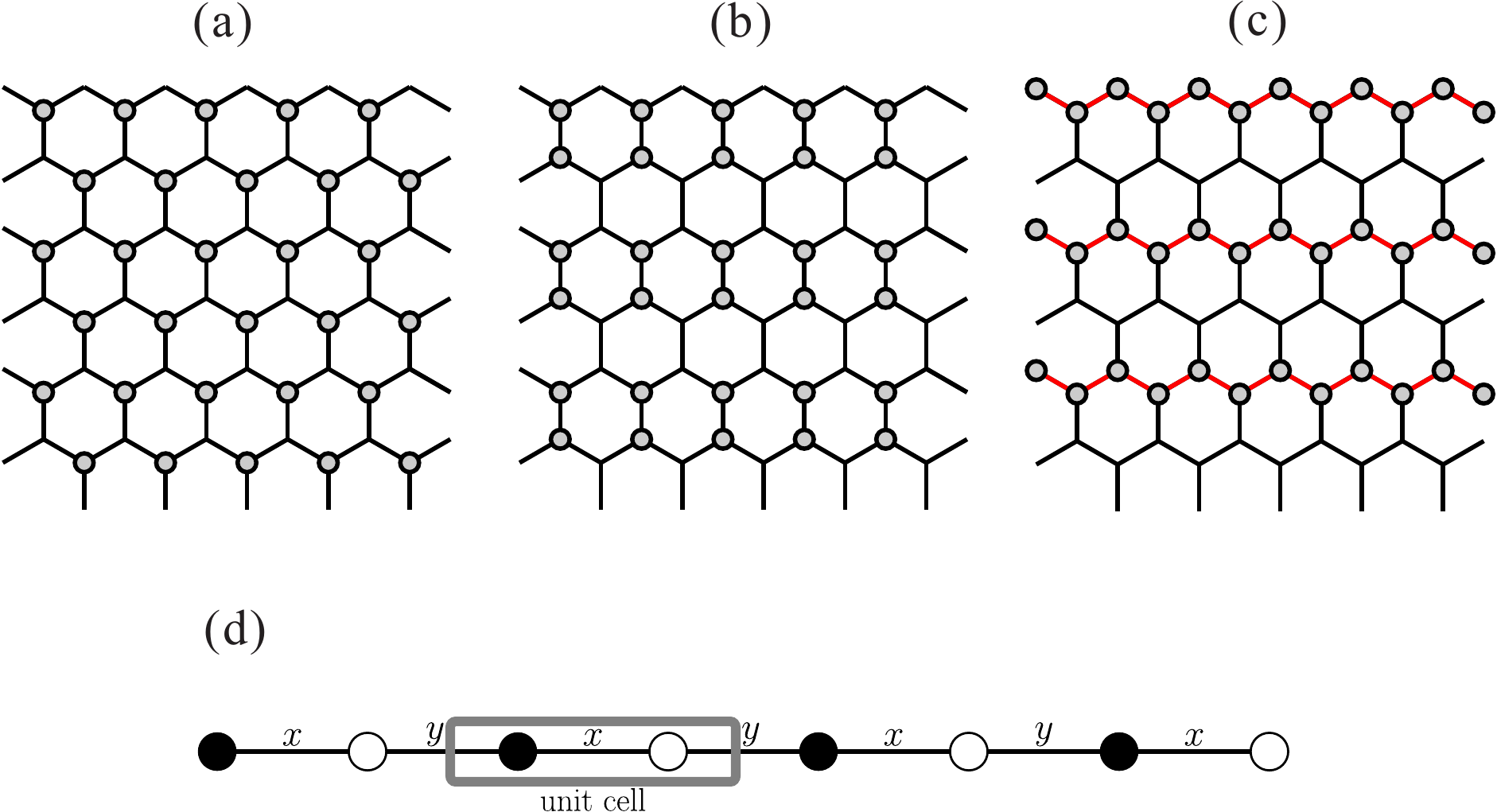}
\caption{
Possible charge ordering patterns of K$_{0.5}$RuCl$_3$; (a) triangular, (b) dimer, and (c) zigzag. (d) Lattice structure of the 1D Heisenberg-Kitaev model. The labels `$x$' and `$y$' indicate the $x$-bond and $y$-bond, respectively (see text).
}
\label{fig_lattice}
\end{figure}

The insulating feature of K$_{0.5}$RuCl$_3$ with a sharp peak at $\sim 0.9$ eV in the EELS spectrum  can be explained by charge ordering associated with a superlattice formation rather than a random distribution of $K^+$ ions (Knupfer M., private communications). The superlattice structures are simply constructed by replacing half of the $j_{\rm eff}=1/2$ pseudospins with nonmagnetic sites in the original hexagonal cluster. In practice, there are three possibilities for the charge ordering pattern, they shown in Fig~\ref{fig_lattice}(a)-(c).  Recently, for exfoliated $\alpha$-RuCl$_3$, a first-order structural phase transition at $\sim 150$ K between low-temperature $C$2/m and high-temperature $P$3$_1$ structures was reported.~\cite{Ziatdinov16} At high temperature, the emergence of charge ordering, originating from anisotropy in the charge distribution along Ru-Cl-Ru hopping pathways, was observed. It suggests a strong tendency to the bond-directional anisotropy along the Ru-Ru axes. In this sense a realization of the zigzag-type charge ordering [Fig~\ref{fig_lattice}(c)] may be most likely. Besides, the zigzag-type charge ordering can gain the largest exchange energy (see below). To fix the detailed charge distribution, experimental observation and/or microscopic analysis including structural distortion by the K-intercalation are required in future.

\subsection*{1D Kitaev-Heisenberg Hamiltonian}

At present, it is widely believed that the magnetic properties of the undoped $\alpha$-RuCl$_3$ are well described by the KH model on a honeycomb lattice. If we assume the zigzag-type charge ordering in Fig~\ref{fig_lattice}(c), the zigzag chains are well separated by the nonmagnetic ions. Then, each chain is considered to be a 1D KH model, which is equivalent to a system obtained by removing the $z$-bonds from the honeycomb-lattice KH model. The Hamiltonian of 1D KH model is written as
\begin{equation}
\mathcal{H}=K\sum_{i=1}^{L/2} (S^x_{2i-1} S^x_{2i}+S_{2i}^y S^y_{2i+1})+J\sum_{i=1}^L \vec{S}_i\cdot\vec{S}_{i+1}
\label{ham}
\end{equation}
for a system with $L$ sites, where $\vec{S}_i=(S_i^x,S_i^y,S_i^z)$ is a spin-$\frac{1}{2}$ operator at site $i$, and the Kitaev and exchange couplings are defined as $K=\sin\phi$ and $J=\cos\phi$ via a phase parameter $\phi$. Throughout the paper, we take $\sqrt{K^2+J^2}=1$ as the energy unit. The system has two kinds of neighboring links and they appear alternately along the chain. Hence, the structural unit cell contains two lattice sites. Hereafter, we call the links ($2i-1$,$2i$) and ($i$,$2i+1$) ``$x$-link'' and ``$y$-link'', respectively. By rewriting Eq.~(\ref{ham}) as
\begin{eqnarray}
\nonumber
\mathcal{H}&=&\frac{2J+K}{4}\sum_{i=1}^L (S^+_i S^-_{i+1}+S^-_i S^+_{i+1})+J\sum_{i=1}^L S^z_i S^z_{i+1} +\frac{K}{4}\sum_i^L(-1)^i(S^+_i S^+_{i+1}+S^-_i S^-_{i+1})\\
\nonumber
&\equiv&{\cal H}_{\rm ex}+{\cal H}_{\rm Ising}+{\cal H}_{\rm dsf},
\label{ham3}
\end{eqnarray}
we can easily notice that a XXZ Heisenberg chain containing exchange (${\cal H}_{\rm ex}$) and Ising (${\cal H}_{\rm Ising}$) terms is disturbed by sign-alternating double-spin-flip (${\cal H}_{\rm dsf}$) fluctuations \cite{Sela14}.

\subsection*{Numerical methods}

We employ the density-matrix renormalization group (DMRG) method, which is one of the most powerful numerical techniques for studying various quasi 1D quantum systems~\cite{white92}. Open boundary conditions are applied unless stated otherwise. This enables us to calculate ground-state and low-lying excited-state energies, as well as static quantities, quite accurately for very large finite-size systems. We are thus allowed to carry out an accurate finite-size-scaling analysis to obtain energies and quantities in the thermodynamic limit $L\to\infty$. We hence study chains with several lengths up to $L=200$ sites for a given $\phi$. Since, in hindsight, the system (\ref{ham}) exhibits only commensurate phases in the ground state and the largest magnetic unit cell contains four lattice sites, its size is taken as $L=4n$ ($n$: integer). For each calculation, we keep  up to $m=1200$ density-matrix eigenstates in the renormalization procedure and extrapolate the calculated quantities to the limit $m \to \infty$ if needed. Since the SU(2) symmetry is broken in system (\ref{ham}) and total $S^z$ is no longer a good quantum number except at $\phi=0$, $\pi$, one may have some difficulty in obtaining accurate results in comparison to usual DMRG calculations. Nevertheless, in this way, the maximum truncation error, i.e., the discarded weight, is less than $1\times10^{-10}$ while the maximum error in the ground-state is less than $1\times10^{-8}$.

For the dynamical properties calculation, we used the Lanczos exact diagonalization (ED) method. To examine the low-energy excitations for  each phase, we calculate the dynamical spin structure factor, defined as
\begin{eqnarray}
\nonumber
S_\gamma(q,\omega) &=& \frac{1}{\pi}{\rm Im} \langle \psi_0 | (S^\gamma_q)^\dagger \frac{1}{\hat{H}+\omega-E_0-{\rm i}\eta} S^\gamma_q | \psi_0 \rangle\\
&=& \sum_\nu |\langle \psi_\nu |S^\gamma_q| \psi_0 \rangle|^2 \delta(\omega-E_\nu+E_0),
\label{spec}
\end{eqnarray}
where $\gamma$ is $z$ or $-(+)$, $| \psi_\nu \rangle$ and $E_\nu$ are the $\nu$-th eingenstate and the eigenenergy  of the system, respectively ($\nu=0$ corresponds to the ground state). Under periodic boundary conditions, the spin operators $S^\gamma_q$ can be precisely defined by
\begin{equation}
S^\gamma_q = \sqrt{\frac{2}{L}}\sum_iS^\gamma_i\exp(iqr_i)
\label{sq}
\end{equation}
where $r_i$ is the position of site $i$ and the sum runs over either $i$ even or $i$ odd sites. They provide the same results. The momentum is taken as $q=\frac{4\pi}{L}n$ ($n=0,\pm1,\dots, \pm\frac{L}{4}$) since the lattice unit cell includes two sites and the number of unit cells is $\frac{L}{2}$ in a system with $L$ sites. We calculate both spectral functions $S_\pm(q,\omega)$ and $S_z(q,\omega)$ as they are different due to  broken SU(2) symmetry except at $\phi=0$ and $\pi$. We study chains with $L=24$, namely, 12 unit cells, by the Lanczos ED method. As shown below,  system (\ref{ham}) contains only commensurate phases with unit cell containing one, two, or four sites. Therefore, a quantitative discussion of the low-lying excitations is possible even within the $L=24$ chain.

\section*{Results}

\subsection*{Ground-state properties} 

\subsubsection*{Quantum phase transitions} 

\begin{figure}[H]
\centering
\includegraphics[width=0.8\linewidth]{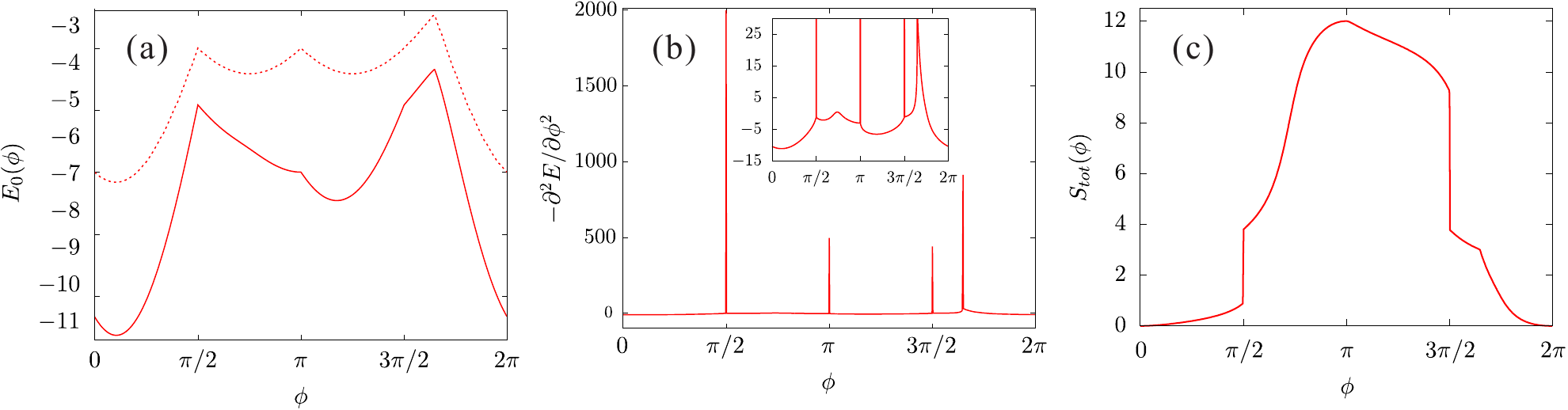}\\
\caption{
(a) Ground-state energy $E_0$, (b) the second derivative of $E_0$ with respect to $\phi$, and (c) total spin $S_{tot}$ as a function of $\phi$, obtaind with a 24-site periodic Kitaev-Heisenberg chain. Dotted line in (a) indicates ground-state energy for the dimer-type charge ordering.
}
\label{fig_energy}
\end{figure}

Lets first look at the ground-state energy and total spin with respect to $\phi$ in order to capture the overall appearance of quantum phase transitions. The results, using a periodic 24-site KH chain, are shown in Fig.~\ref{fig_energy}. We clearly see discontinuities in the first derivative of the ground-state energy at four values: $\phi=\frac{\pi}{2}$, $\pi$, $\frac{3\pi}{2}$, and $\approx1.65\pi$, which indicate the first-order phase transitions. The second derivative $-\partial^2E_0/\partial \phi^2$ is, in a precise sense, continuous except for the above four $\phi$ points, nonetheless,  there exists a distinguishable peak around $\phi \approx 0.65\pi$, which may corresponds to a second-order (or continuous) phase transition. Furthermore, as shown below, we find another phase transitions at $\phi=0$. Therefore, we suggest that the simple 1D model (\ref{ham}) exhibits a variety of phases including six quantum phase transitions. It will be confirmed by studying various corresponding order parameters or spin-spin correlation functions. We also confirm that the ground-state energy of the 1D KH chain is always lower than that given by the dimer-type charge ordering [see Fig.~\ref{fig_energy}(a)].

\begin{figure}[!t]
\centering
\includegraphics[width=0.4\linewidth]{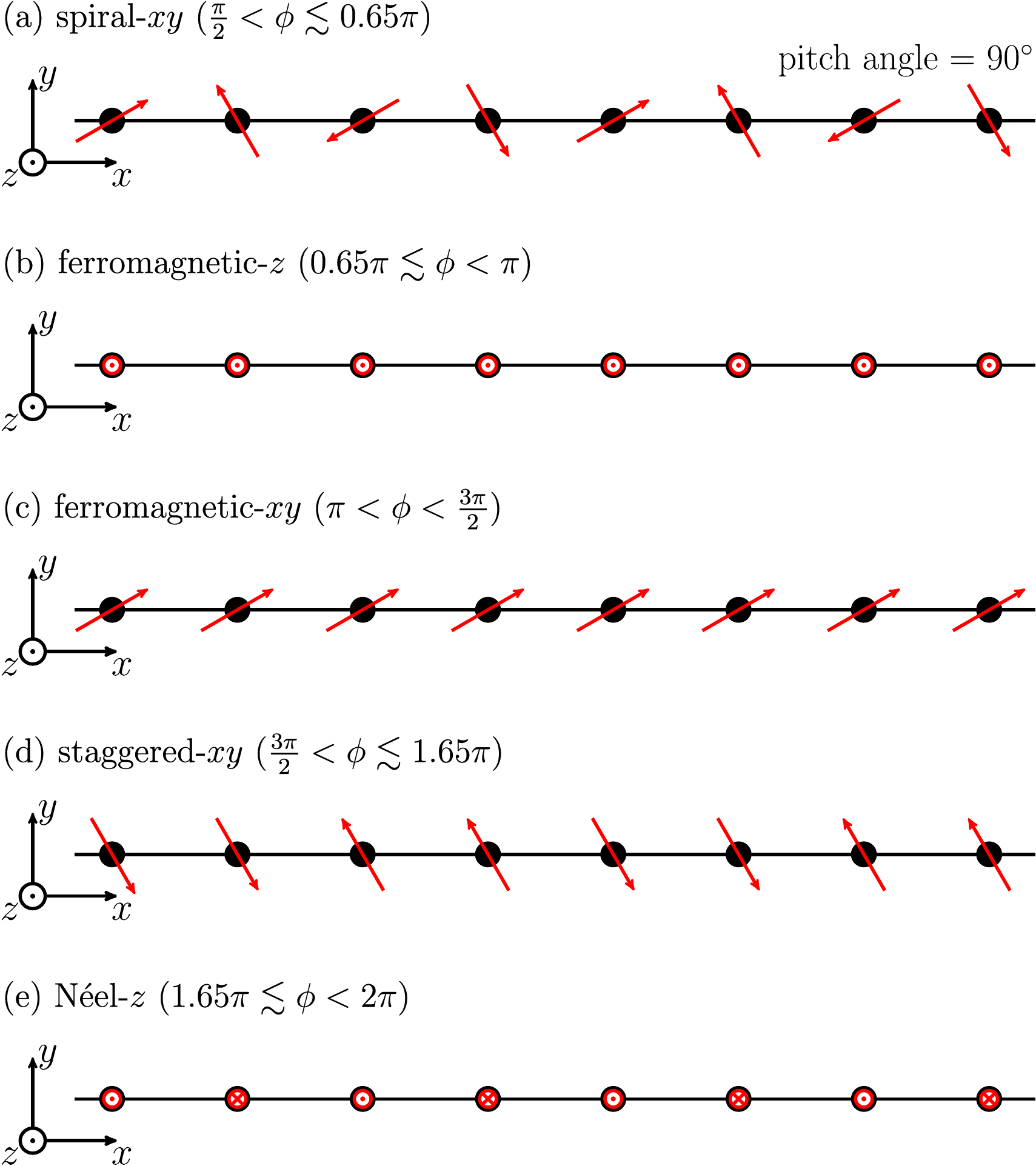}\\
\caption{
Schematic pictures of five states realized in the 1D KH model, except the TLL state at $0<\phi<\frac{\pi}{2}$. In the phases (a)(c)(d), the spins lie mostly on the xy-plane and these states are rotational invariant around the $z$-axis. In phases (b)(e), the spins almost align along the $z$-direction and a state having opposite spin directions is degenerate.
}
\label{fig_states}
\end{figure}

\subsubsection*{Ferromagnetic-$xy$ phase ($\pi<\phi<\frac{3\pi}{2}$)}

Since both $K$ and $J$ are ferromagnetic (FM), a long-range FM ordered state is naively expected in the range $\pi < \phi < \frac{3\pi}{2}$. At the spin isotropic point $\phi=\pi$, the spins can align along any arbitrary spatial direction due to SU(2) spin rotation invariance and the total spin takes the maximum value $\frac{S_{tot}}{L}=\frac{1}{2}$ [see Fig.~\ref{fig_energy}(c)]. Away from the isotropic point, the SU(2) symmetry is broken to U(1) and the configurations for higher $|S^z|$ sectors are projected out [see Fig.~\ref{fig_states}(c)]. When $\phi$ is still close to $\pi$ ($-J \gg -K > 0$), the ground state is approximately expressed as
\begin{equation}
|\Psi_0\rangle=\frac{1}{\sqrt{\cal N}}\sum_m |\psi_m \rangle,
\label{fm1}
\end{equation}
where $|\psi_m \rangle$ is a basis in real space, namely each site state is represented by either spin-up ($\uparrow$) or spin-down ($\downarrow$). The basis is here restricted to a $S^z_{\rm tot}=\sum_{i=1}^L \langle S^z_i \rangle=0$ subspace, so that $m$ is summed over all possible combinations of spin configuration with $\frac{L}{2}$ up and $\frac{L}{2}$ down spins in a $L$ lattice sites. ${\cal N}$ is the total number of the spin configurations, i.e., $\frac{L!}{(L/2)!(L/2)!}$. The wave function (\ref{fm1}) becomes exact in the isotropic spin limit $\phi=\pi+$. Accordingly, the spin-spin correlations have long-range FM ordering for all three spin components: $\langle S^x_i S^x_j \rangle = \langle S^y_i S^y_j \rangle = \frac{1}{6}$ and $\langle S^z_i S^z_j \rangle = -\frac{1}{12}$ for any $i\neq j$. Taking the spin isotropic Hamiltonian at $\phi=\pi+$ as an unperturbed one, the unperturbed ground state is given by Eq.~(\ref{fm1}). When $\phi-\pi \ll 1$, the perturbed Hamiltonian can be written as ${\cal H}^\prime\approx\frac{\phi-\pi}{4}\sum_i(S_i^+S_{i+1}^-+S_i^-S_{i+1}^+)$ and the lowest-order energy correction is $E^\prime=\frac{\phi-\pi}{4}$. Therefore, with increasing $\phi$ from $\pi$, the antiferromagnetic (AFM) fluctuations increase and the long-range FM ordering is weakened. Nonetheless, the correlations $\langle S^x_i S^x_j \rangle$ and $\langle S^y_i S^y_j \rangle$ retain the same asymptotic behaviors indicating the long-range FM ordering until $\phi=\frac{3\pi}{2}$, characterized by the saturation to a finite negative value in the limit $|i-j| \to \infty$; whereas, $\langle S^z_i S^z_j \rangle$ decays in a power law with $|i-j|$. This is because the AFM fluctuations are mainly introduced along the $z$-direction. It is confirmed by a slow decrease of the total spin as a function of $\phi$, as seen in Fig.~\ref{fig_energy}(c). We thus call this state FM-$xy$ state. A collapse of the long-range FM ordering is detected by a drop-off of the total spin at $\phi=\frac{3\pi}{2}$, suggesting  a first-order transition. 

\subsubsection*{Ferromagnetic-$z$ phase ($0.65\pi\lesssim\phi<\pi$)}

\begin{figure}[b]
\centering
\vspace*{-5mm}\includegraphics[width=0.60\linewidth]{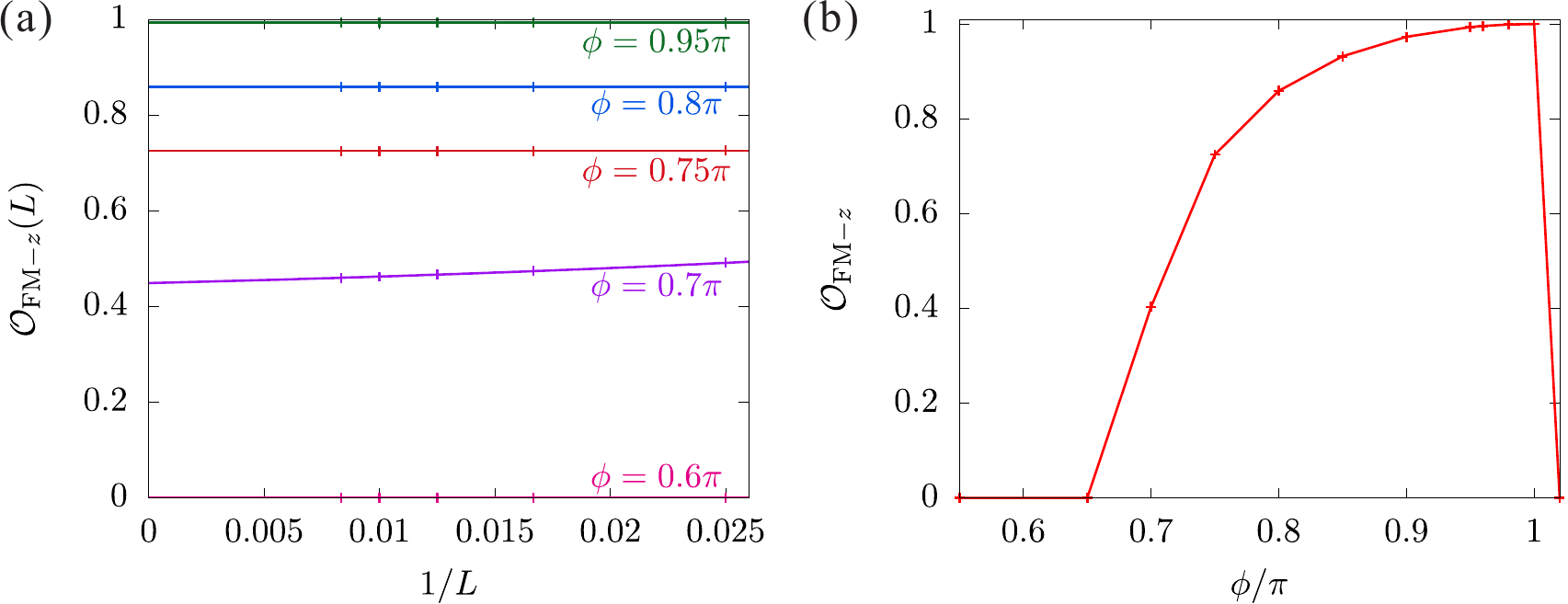}
\caption{
(a) Finite-size scaling analyses of the FM-$z$ order parameter with polynomial fitting functions. (b) Extrapolated values of the FM-$z$ order parameter ${\cal O}_{{\rm FM}-z}$ to the thermodynamic limit $L\to\infty$ as a function of $\phi$.
}
\label{SzOP}
\end{figure}

A long-range FM ordered state stabilizes also at $\frac{3\pi}{4}<\phi<\pi$, where all the exchange tensors are FM since $J+K$ is negative despite positive $K$. However, in contrast to the FM-$xy$ state, spin alignment along the $z$-direction is favored because of the easy-axis-XXZ-like interactions $|\frac{2J+K}{2}|<|J|$. Therefore, the most dominant spin configurations are given by the highest $|S^z|$ sectors, i.e., the ground state is expressed as
\begin{equation}
|\Psi_0\rangle=\frac{1}{\sqrt{2}}(|\Uparrow \rangle+|\Downarrow \rangle),
\label{fm2}
\end{equation}
with $|\Uparrow \rangle=|\cdots \uparrow \uparrow \uparrow \uparrow \uparrow \uparrow \cdots \rangle$ and 
$|\Downarrow \rangle=|\cdots \downarrow \downarrow \downarrow \downarrow \downarrow \downarrow \cdots \rangle$. This wave function is exact in the isotropic limit of $\phi=\pi-$. We call this type of long-range FM ordering FM-$z$ state. Let us take Eq.~(\ref{fm2}) as the unperturbed ground state. For $\pi-\phi \ll 1$, the perturbed Hamiltonian can be written as ${\cal H}^\prime\approx\frac{(\pi-\phi)^2}{2}\sum_i S_i^zS_{i+1}^z$. This is an Ising-like AFM correlation and it clearly disturbs the FM-$z$ state. 
However, the lowest-order correction to the ground-state energy is $E^\prime=\frac{(\pi-\phi)^2}{8}$. It means that the perturbation acts only gradually with being away from the isotropic point $\phi=\pi$. As a result, the total spin is not that much reduced around $\phi=\pi$ in the FM-$z$ phase. To estimate the lower bound of the FM-$z$ phase, we shall define the FM-$z$ order parameter. A state with long-range FM ordering is a state with broken spin symmetry along the $z$-direction; macroscopically, there are two degenerate ground states $|\Psi_0\rangle\approx|\Uparrow \rangle$ and $|\Psi_0\rangle\approx|\Downarrow \rangle$. Applying open boundary conditions, one of the two ground states is picked imposing initial conditions on the calculation. The long-range ordered state is thus directly observable as a symmetry-broken state in our DMRG calculations. The order parameter is defined as
\begin{equation}
{\cal O}_{{\rm FM}-z}=2\lim_{L \to \infty} \langle S^z_{L/2} \rangle
\label{OFMz}
\end{equation}
The finite-size scaling analysis is performed. The extrapolated values in the thermodynamic limit are shown in Fig.~\ref{SzOP}. Notably, the FM-$z$ state survives even at $\phi<\frac{3\pi}{4}$, where a part of the interactions is AFM, i.e., $J+K>0$. Nevertheless, the long-ranged FM order is drastically suppressed by the AFM fluctuations at $\phi<\frac{3\pi}{4}$ and completely destroyed at $\phi\approx0.65\pi$. The order parameter has no jump at the transition point, suggesting a second-order phase transition.

\subsubsection*{Spiral-$xy$ phase ($\frac{\pi}{2}<\phi\lesssim0.65\pi$)} 

\begin{figure}[!t]
\centering
\includegraphics[width=0.80\linewidth]{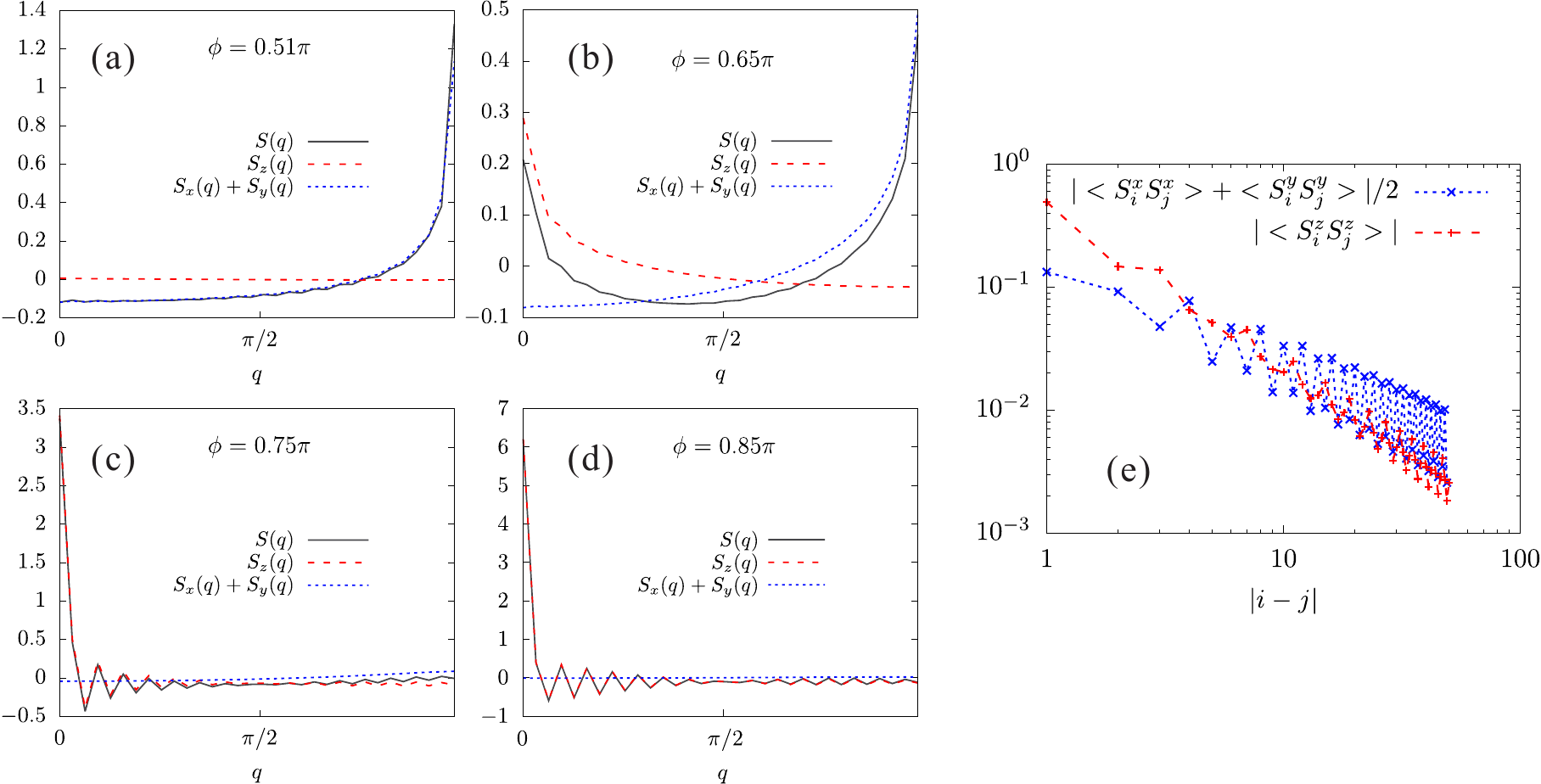}
\caption{
Static spin structure factors for (a) $\phi=0.51\pi$, (b) $\phi=0.6\pi$, (c) $\phi=0.65\pi$, and (d) $\phi=0.75\pi$. (e) Log-log plot of the spin-spin correlation functions as a function of distance at $\phi=0.6\pi$.
}
\label{Sq}
\end{figure}

As shown above, the FM-$z$ phase remains down to $\phi\approx0.65\pi$. Then, we ask which kind of phase comes next at $\frac{\pi}{2}<\phi\lesssim0.65\pi$. In this range, $K$ is AFM, $J$ is FM, and $K \gg -J$. On the $x$-links, the $x$-components of spins tend to be antiparallel due to the strong AFM interaction along the $x$ direction; whereas, their $y$-components tend to be parallel due to FM $J$ term. One can think about the $y$-links in the same way. Eventually, the spins lie on the $xy$ plane and rotate by 90$^\circ$ from one site to the next. The magnetic unit cell is twice as large as the structural unit cell, i.e., including four lattice sites [see Fig.~\ref{fig_states}(a)]. We call this state spiral-$xy$  state. To confirm this magnetic structure, we calculate the static spin structure factor, defined as
\begin{equation}
S^\gamma(q)=\frac{2}{L}\sum_{k,l}\langle S_k^\gamma S_l^\gamma \rangle e^{-iq(k-l)},
\label{Sq_def}
\end{equation}
where the length of the structural unit cell, i.e., two lattice spacings, is taken to be unity. The DMRG results with $L=60$ cluster are shown in Fig.~\ref{Sq}. Near the AFM Kitaev point $\phi=\frac{\pi}{2}$, $S^x(q)+S^y(q)$ has a large $q=\pi$ peak indicating a periodicity of four lattice sites on the $xy$-plain. On the other hand, $S_z(q)$ is almost zero for all $q$ since the FM $z$ interaction $J$ is tiny compared to the dominant AFM $x$ or $y$ interactions $J+K$. With increasing $\phi$, the $q=\pi$ peak becomes lower and a $q=0$ peak in $S^z(q)$ develops. Basically, the spins are tilted in one direction along the $z$-axis with keeping the periodicity on the $xy$-plain. Those peak heights are reversed around the transition point $\phi=0.65\pi$ from the spiral-$xy$ to the FM-$z$ phases. We note that the height of the $q=0$ peak in $S^z(q)$ coincides with the FM-$z$ order parameter. As shown in Fig.~\ref{Sq}(e), the spin-spin correlations decay as a power law for all the spin components in the spiral-$xy$ phase. It means that the $q=\pi$ peak in $S^x(q)+S^y(q)$ disappears in the thermodynamic limit: the spiral-$xy$ structure is not long-range ordered. 

\subsubsection*{N\'eel-$z$ ordered phase ($1.65\pi\lesssim\phi<2\pi$)} 

\begin{figure}[!t]
\centering
\vspace*{-5mm}\includegraphics[width=0.80\linewidth]{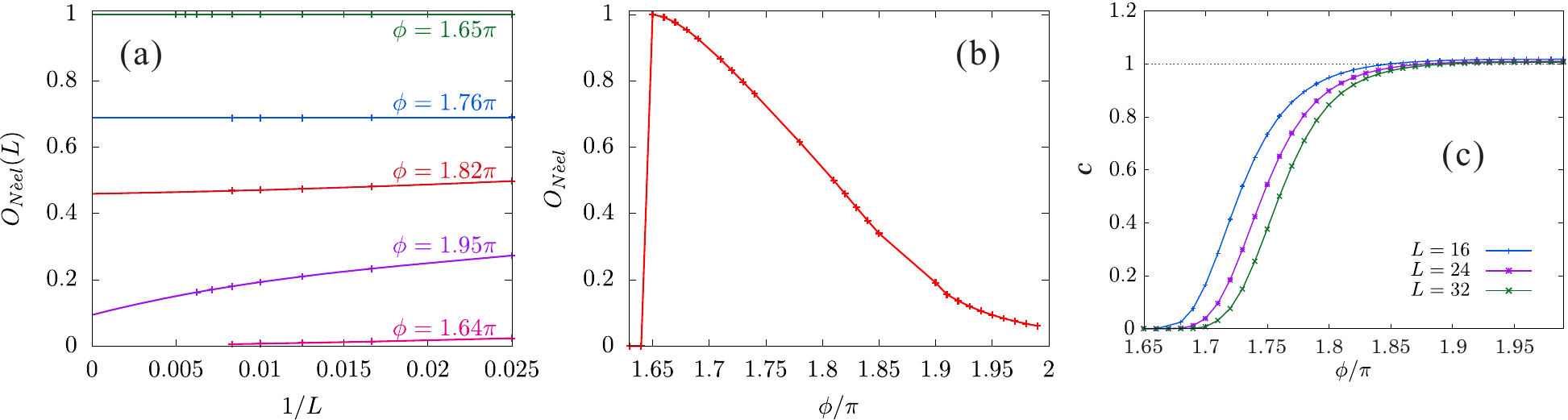}\\
\caption{
(a) Finite-size scaling analyses for the N\'eel-$z$ order parameter ${\cal O}_{\textrm{N\'eel}}$ with polynomial fitting functions. (b) The extrapolated values of ${\cal O}_{\textrm{N\'eel}}$ to the themodynamic limit as a function of $\phi$. (c) Central charge, calculated with periodic chains,  as a function of $\phi$ for several chain lengths.
}
\label{NeelOP}
\end{figure}

Next, we turn to the parameter region $\frac{3\pi}{2}<\phi<2\pi$ ($J>0$, $K<0$). We start from the SU(2) symmetric limit $\phi=2\pi$, where the system is the original 1D AFM Heisenberg model. The ground-state wave function can be exactly obtained and it is know to have $S_{\rm tot}=0$. So, the first perturbative correction is given by ${\cal H}^\prime\approx(\phi-2\pi)\sum_i(S^+_iS^-_{i+1}+S^-_iS^+_{i+1})$, which provides an easy-axis anisotropy to the system. Therefore, the possibility of a continuous N\'eel-$z$ order transition is conceived by analogy to the easy-axis anisotropic XXZ chain. For small $2\pi-\phi$, the magnetization is expect to grow gradually with
\begin{equation}
M\simeq\frac{\pi}{2(2\pi-\phi)}\exp \left\{-\frac{\pi^2}{2\sqrt{2(2\pi-\phi)}}\right\}.
\label{mag}
\end{equation}

Let us confirm it numerically. In the long-range N\'eel-$z$ ordered state, the translational symmetry is broken in a finite system due to the Friedel oscillation under the open boundary conditions, so that the N\'eel-$z$ state can be directly observed by extracting one of the degenerate states, like in the FM-$z$ state. Generally, the Friedel oscillations in the center of the system decay as a function of the system length. If the amplitude at the center of the system persists for arbitrary system lengths, it corresponds to a long-range ordering. Here, we are interested in the formation of alternating spin flip along the $z$-direction. Thus, the N\'eel-$z$ order parameter is defined as
\begin{equation}
{\cal O}_{\textrm{N\'eel}}=\lim_{L\to\infty}|\langle S_{L/2}^z \rangle - \langle S_{L/2+1}^z \rangle|.
\end{equation}
This quantity is equivalent to the magnetization $M$ in the thermodynamic limit. The finite-size scaling analyses of ${\cal O}_{\textrm{N\'eel}}$ was performed using the results for systems with up to $L=120$ and the extrapolated values to the thermodynamic limit were obtained. They are shown in Fig.~\ref{NeelOP}. We indeed see a slow increase of ${\cal O}_{\textrm{N\'eel}}$ near the SU(2) symmetric limit $\phi=2\pi$. Further with decreasing $\phi$, the order parameter develops up to ${\cal O}_{\textrm{N\'eel}}=1$ and drops down to $0$ at $\phi\approx1.65\pi$.

In fact, this lower $\phi$-boundary of the N\'eel-$z$ ordered phase can be estimated analytically. At $\phi=\tan^{-1}(-2)\approx1.6476\pi$, the exchange term ${\cal H}_{\rm ex}$ disappears in the Hamiltonian (\ref{ham}) and the system is just written as a sum of double-spin-flip and Ising parts
\begin{eqnarray}
\nonumber
{\cal H}&=&{\cal H}_{\rm dsf}+{\cal H}_{\rm Ising}\\
&=&\frac{K}{4}\sum_i^L (-1)^i (S^+_iS^+_{i+1}+S^-_iS^-_{i+1})+\frac{J}{4}\sum_i^L S^z_iS^z_{i+1}.
\label{ham2}
\end{eqnarray}
Each of the partition Hamiltonians ${\cal H}_{\rm dsf}$ and ${\cal H}_{\rm Ising}$ is exactly solvable. For ${\cal H}_{\rm dsf}$, the system is regarded as noninteracting fermions with ``pair hopping`' and the ground-state wave function is
\begin{equation}
|\Psi_0\rangle=\frac{1}{\sqrt{\cal N}}\sum_{\cal C}\prod_{i({\cal C}),\gamma({\cal C})} (-1)^{i-1}S_i^\gamma S_{i+1}^\gamma \ket\Uparrow,
\label{XY1}
\end{equation}
where ${\cal C}$ is summed over all possible spin configurations created by the double spin flips starting from $\ket \Uparrow$ or $\ket \Downarrow$ (${\cal N}=_LC_{L/2}$), $i$ and $\gamma$(=+ or -) are taken to create the configuration ${\cal C}$. While, for ${\cal H}_{\rm Ising}$, the system is a simple Ising one and the ground-state wave function is
\begin{equation}
|\Psi_0\rangle=\frac{1}{\sqrt{2}}\sum_\sigma\prod_{i=1}^{L/2}c_{2i-1,\sigma}^\dagger c_{2i,\bar{\sigma}}^\dagger\ket0,
\label{Neel1}
\end{equation}
where $\ket0$ is the vacuum state. The ground-state wave functions (\ref{XY1}) and (\ref{Neel1}) are orthogonal since they do not share the same spin configurations. It means the ground state is two-fold degenerate at the critical point $\phi=\tan^{-1}(-2)$. The ground-state energy is $E_0=-\frac{L\cos \phi}{4}$. This degeneracy is lifted away from $\phi=\tan^{-1}(-2)$. Eq.~(\ref{Neel1}) is the ground state for larger-$\phi$ side, namely, in the N\'eel-$z$ phase; while, Eq.~(\ref{XY1}) for lower-$\phi$ side (see Sec.~\ref{StXY}). This is consistent with the fact that ${\cal O}_{\textrm{N\'eel}}$ reaches exactly 0.5 at $\phi=\tan^{-1}(-2)$, as shown in Fig.~\ref{NeelOP}(b). Therefore, the ground-state wave function is completely changed at $\phi=\tan^{-1}(-2)$ and it clearly suggests a first-order transition. This is also confirmed by the jump of ${\cal O}_{\textrm{N\'eel}}$ 
as well as by the singularity in $-\partial^2E_0/\partial \phi^2$ at $\phi\approx1.65\pi$.

\subsubsection*{Staggered-$xy$ ordered phase ($\frac{3\pi}{2}<\phi\lesssim1.65\pi$)} 
\label{StXY}

\begin{figure}[!t]
\centering
\includegraphics[width=0.80\linewidth]{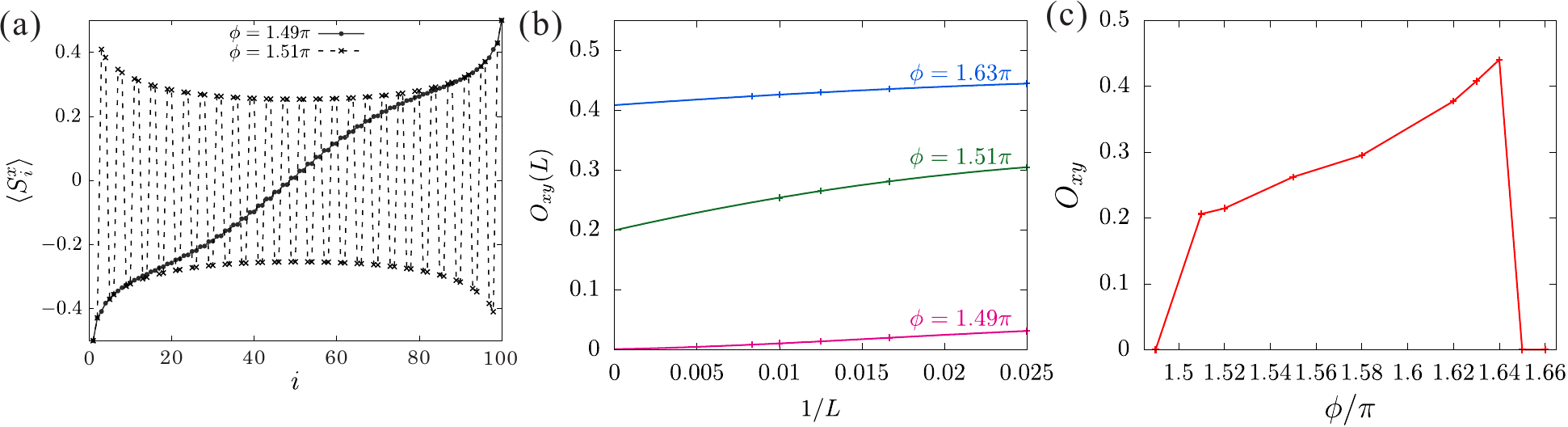}\\
\caption{
(a) The $x$-component of local spin, where opposite magnetic fields $\pm 10$ are applied at either end of the system. (b) Finite-size scaling analyses for the staggered-$xy$ order parameter with polynomial fitting functions. (c) The extrapolated values of ${\cal O}_{xy}$ to the themodynamic limit as a function of $\phi$.
}
\label{fig_XYOP}
\end{figure}

Let us then consider a region at $\frac{3\pi}{2}<\phi\lesssim1.65\pi$, where the signs of Heisenberg and Kitaev terms counterchange  from the spiral-$xy$ state. This may mean the local spin structure is similar to the spiral-$xy$ state. As discussed above, the system has a first-order transition at $\phi=\tan^{-1}(-2)\approx1.65\pi$. In the lower-$\phi$ vicinity ($\phi\lesssim1.65$) the ground state is exactly given by Eq.~(\ref{XY1}), which derives asymptotic behaviors of the spin-spin correlations: $\langle S^x_0 S^x_j \rangle=\frac{\alpha^2}{4\sqrt{2}}\cos\frac{\pi}{2}[(j+\frac{1}{2})]$, $\langle S^y_0 S^y_j \rangle=\frac{\beta^2}{4\sqrt{2}}\cos\frac{\pi}{2}[(j-\frac{1}{2})]$, $\langle S^z_0 S^z_j \rangle=0$, and $\langle \vec{S}_0 \vec{S}_j \rangle=\frac{1}{4}\cos\left(\frac{\pi}{2}j\right)+\frac{\alpha^2-\beta^2}{4}\sin\left(\frac{\pi}{2}j\right)$. Since the system is rotation invariant around the $z$-axis, the coefficients $\alpha$ and $\beta$ can take arbitrary real numbers under a condition $\alpha^2+\beta^2=1$. As illustrated in Fig.~\ref{fig_states}(d), all the spins lie on the $xy$-plane and the magnetic unit cell contains four lattice sites as in the spiral-$xy$ state. However, the crucial difference from the spiral-$xy$ state is that this $xy$ state exhibits a long-range ordering, which is clearly indicated by the correlation functions. We call this state staggered-$xy$ state, since it resembles a N\'eel-like state with a period of four sites.

We then investigate the $\phi$-dependence of this $xy$ state. Applying a staggered field along the $x$-direction on both open system edge sites, the presence or absence of the long-range staggered-$xy$ order can be determined by studying the decay of the $x$-component of local spin $\langle S^x_i \rangle$ as the Friedel oscillation. Thus, the staggered-$xy$ order parameter is defined at the center of the system as
\begin{equation}
{\cal O}_{xy}=\frac{1}{2}\lim_{L\to\infty}|\langle S_{L/2}^x \rangle - \langle S_{L/2+1}^x \rangle|.
\label{XYOP}
\end{equation}
In the upper and lower vicinity of $\phi\approx1.65\pi$, ${\cal O}_{xy}=0$ and $\frac{1}{2}$ are obtained from Eqs.~(\ref{XY1}) and (\ref{Neel1}), respectively. For the other $\phi$ values the profile of the $x$-component of local spin and the finite-size scaling of the order parameter are shown in Fig.~\ref{fig_XYOP}(a)(b). The extrapolated values of ${\cal O}_{xy}$ to the thermodynamic limit are plotted in Fig.~\ref{fig_XYOP}(c). We can see that the order parameter jumps at the both phase boundaries suggesting first-order transitions. The collapse of the staggered-$xy$ ordering at the lower boundary $\phi=\frac{3\pi}{2}$ is clearly confirmed by the profile of $\langle S^x_i \rangle$, as shown in Fig.~\ref{fig_XYOP}(a).

\begin{figure}[!t]
\centering
\vspace*{-1mm}\includegraphics[width=0.4\linewidth]{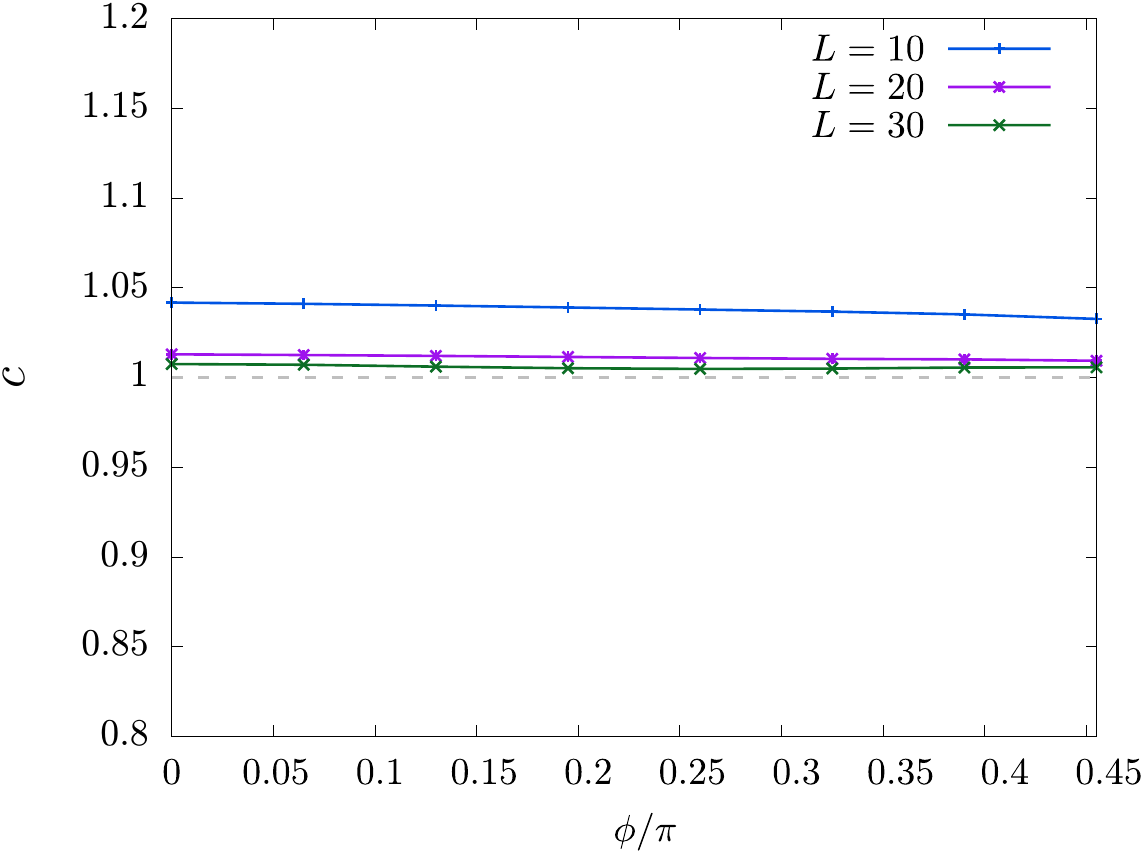}\\
\caption{
Central charge, calculated with periodic $L=10$, $20$, and $30$ chains,  as a function of $\phi$ .
}
\label{TLLcc}
\end{figure}

\subsubsection*{Tomonaga-Luttinger-liquid phase ($0 \le \phi <\frac{\pi}{2}$)} 

The remaining region is $0 \le \phi <\frac{\pi}{2}$. At $\phi=0$ the system is equivalent to the spin-isotropic AFM Heisenberg chain, which is a gapless spin liquid. It is known that the low-energy physics is described by a Tomonaga-Luttinger (TL) model with a boson field, which is equivalent to the unity central charge ($c=1$) conformal field theory (CFT)~\cite{Gogolin98}. Let us now consider what happens when we move away from $\phi=0$. At $0<\phi<\frac{\pi}{2}$ the deviating interactions from the spin-isotropic Hamiltonian are written as
\begin{equation}
\mathcal{H}_{\rm aniso}=\frac{K}{4}\sum_{i=1}^L [S^+_i S^-_{i+1}+S_{i}^- S^+_{i+1}+(-1)^i(S^+_i S^+_{i+1}+S_{i}^- S^-_{i+1})].
\label{hamTL}
\end{equation}
Obviously, these interactions cannot produce any explicit magnetic order, since they only enhance the quantum fluctuations. However, it is a nontrivial question whether the $c=1$ CFT is conserved. To examine it, we directly calculate the central charge, which can be obtained from the von Neumann entanglement entropy in the DMRG procedure as follows: Let us consider a quantum 1D periodic system with length $L$. The von Neumann entanglement entropy of its subsystem with length $l$ is given as $S_L(l)=-{\rm Tr}_l \rho_l \log \rho_l$, where $\rho_l={\rm Tr}_{L-l}\rho$ is the reduced density matrix of the subsystem and $\rho$ is the full density matrix of the whole system. Using the CFT, the entropy of the subsystem with length $l$ for a fixed system length $L$ has been derived:~\cite{Affleck91,Holzhey94,Calabrese04}
\begin{eqnarray}
S_L(l)=\frac{c}{3}\ln\left[\frac{L}{\pi}\sin\left(\frac{\pi l}{L}\right)\right]+s_1
\label{entropy}
\end{eqnarray}
where $s_1$ is a non-universal constant. In the DMRG calculations this quantity can be accurately estimated by the second derivative of Eq.~(\ref{entropy}) with respect to $l$, namely,
\begin{eqnarray}
c=-\frac{3L^2}{\pi^2}\frac{\partial^2S_L(l)}{\partial l^2}\bigg|_{l=\frac{L}{2}}.
\label{cc}
\end{eqnarray}
The results for periodic chains with L=10, 20, and 30 are plotted in Fig.~\ref{TLLcc}. We see a quick convergence to $c=1$ with increasing system size in the whole range of $0 \le \phi <\frac{\pi}{2}$. Hence, we confirm the system remains in the TL liquid state as far as both $J$ and $K$ are AFM. Similarly, we studied the central charge in the N\'eel-$z$ phase. The results for periodic chains with $L=16$, $24$, and $32$ are plotted in Fig.~\ref{NeelOP}(c). Although $c=0$ should be retained in the N\'eel-$z$ phase, we see a very slow converge to $c=0$ with increasing $L$ near $\phi=2\pi$, reflecting the tiny magnetization.

\subsubsection*{Kitaev points} 
\label{Kitaevpoints}

As discussed above, the Kitaev points are singular, not adiabatically connected to the neighboring phases. At these points $\phi=\pm\frac{\pi}{2}$, the Heisenberg interaction $J$ vanishes in our Hamiltonian Eq.~(\ref{ham}) and it is reduced to
\begin{equation}
	\mathcal{H}=\pm K\sum_{x-links} S^x_i S^x_{i+1}\pm K\sum_{y-links} S_i^y S^y_{i+1}
	\label{eq:KHam}
\end{equation}
In order to make the notation easier, we here focus on the case of $\phi=-\pi/2$. Note that the case of $\phi=\frac{\pi}{2}$ can be similarly considered. It is convenient to fermionize the Hamiltonian (\ref{eq:KHam}). By applying the Jordan-Wigner transformation, we obtain 
\begin{eqnarray}
	\mathcal{H}= \sum_{x-links} (c_{{\rm b},i}^\dagger - c_{{\rm b},i}) (c_{{\rm w},i}^\dagger+c_{{\rm w},i})\\ - \sum_{y-links} (c_{{\rm w},i}^\dagger + c_{{\rm w},i}) (c_{{\rm b},i+1}^\dagger - c_{{\rm b},i+1}),
	\label{eq:FermHam}
\end{eqnarray}
where $c_{{\rm b},i}$ and $c_{{\rm w},i}$ are the fermion annihilation operators at the black and white labeling sites in the $i$-th unit cell, illustrated in Fig.~\ref{fig_lattice}(d), respectively. Generally, it is possible to write fermion operators in terms of Majorana fermions defined as:
\begin{align}
	c_{{\rm b},i}^\dagger &=\frac{b_{1,i} - i b_{2,i}}{2} \label{eq:Ferm1}\\
	c_{{\rm w},i}^\dagger &=\frac{w_{1,i} - i w_{2,i}}{2} \label{eq:Ferm2}
\end{align}
With these operators Eq.~(\ref{eq:FermHam}) is transformed to
\begin{equation}
	\mathcal{H}=-i\frac{K}{4} \sum_{x-links} b_{2,i} w_{1,i} + i\frac{K}{4} \sum_{y-links} w_{1,i} b_{2,i+1}.
	\label{eq:MFHam}
\end{equation}
We immediately notice that the Majoranas $b_1,\, w_2$ are not included in the Hamiltonian (\ref{eq:MFHam}). We then define new, non local, fermion operators as
\begin{align}
	d_i^\dagger &=\frac{w_{1,i}-ib_{2,i}}{2} \label{eq:NewFerm1} \\
	d_i &= \frac{w_{1,i}+i b_{2,i}}{2} \label{eq:NewFerm2},
\end{align}
where the indices $i$ run over the unit cells instead of over the lattice sites [see Fig.~(\ref{fig_lattice})(d)]. Using these operators, \eqref{eq:MFHam} can be expressed as
\begin{align}
	\mathcal{H}=\frac{K}{2}\sum_i d_i^\dagger d_i -\frac{K}{4}\sum_i (d_i^\dagger d_{i+1} + d_i d_{i+1} + \text{h.c.}).
\end{align}
This describes a $p$-wave paired superconductor~\cite{Kitaev01}. The exact solution for the ground state is easily obtained by a Fourier transformation:
\begin{equation}
	\mathcal{H}=\frac{1}{2}\sum_q \left[\varepsilon_q d^\dagger_q d_q+i\frac{\Delta_q}{2}(d^\dagger_q d^\dagger_{-q}+\text{h.c.})\right]
\end{equation}
where $\varepsilon_q=-\frac{K}{2}(\cos q+1)$ and $\Delta_q=\frac{K}{2}\sin q$.
Using a Bogoliubov transformation, this can be diagonalized and the quasiparticle excitation is given by
\begin{equation}
E_q=\pm \frac{1}{4}\sqrt{\varepsilon_q^2+\Delta_q^2}.
\end{equation}
Thus, we can confirm the system is still in the gapless region. The ground-state energy is calculated as
\begin{equation}
\frac{E_0}{L}=\frac{1}{L}\sum_q (-E_q)=\frac{1}{2\pi}\int_{-\pi}^\pi dq (-E_q)=-\frac{K}{2\pi}.
\end{equation}

To consider the meaning of the Majoranas $b_1,\, w_2$ which do not appear explicitly in the Hamiltonian (\ref{eq:MFHam}), we take a possible recombination of them into new fermion operators:
\begin{eqnarray}
	\widetilde{d}_i^{\,\dagger}&=&\frac{b_{1,i}-iw_{2,i}}{2}
        \label{recomb1}\\
	\widetilde{d_i} &=&\frac{b_{1,i}+iw_{2,i}}{2}
        \label{recomb2}
\end{eqnarray}
We can now trace these back to the initial spin operators. Using Eqs. \eqref{eq:Ferm1} and \eqref{eq:Ferm2}, we get
\begin{equation}
	\widetilde{d}_i^{\, \dagger} = \frac{1}{2} \left( c_{b,i}^\dagger + c_{b,i} + c_{w,i}^\dagger -c_{w,i}\right).
	\label{tddag1}
\end{equation}
Moreover, the inverse of Jordan-Wigner transformation used to obtain Eq.~(\ref{eq:FermHam}) derives
\begin{align}
	\widetilde {d}_i^{\,\dagger} = \prod_{k=1}^{i-1}\prod_{k'=b,w}(S^z_{k,k'}) ( S_{b,i}^x -i S_{b,i}^z S_{w,i}^y)
	\label{tddag2}
\end{align}
Thus, the Majorana operators can be expressed by spin operators, like
\begin{eqnarray}
	b_{1,i} &=& \widetilde d_i^\dagger + \widetilde d_i = 2\prod_{k=1}^{i-1}(-S_{k,k'}^z)S_{b,i}^x
	\label{bwspin1}\\
	w_{2,i} &=& i(\widetilde d_i^\dagger - \widetilde d_i) = 2\prod_{k'=w,b} (-S_{k',k}^z)(-S_{b,i}^z)(-S_{w,i}^y)
	\label{bwspin2}
\end{eqnarray}
We now confirm that  the system has a free spin per unit cell at the Kitaev points. Therefore, the dimensionality of the ground-state manifold, or number of zero Majorana modes, is $2^{\frac{L}{2}-1}$ for periodic chain and $2^{\frac{L}{2}}$ for open chain. Some more details are given in the Appendix.

\subsection*{Phase diagram} 

\begin{figure}[!t]
\centering
\includegraphics[width=0.60\linewidth]{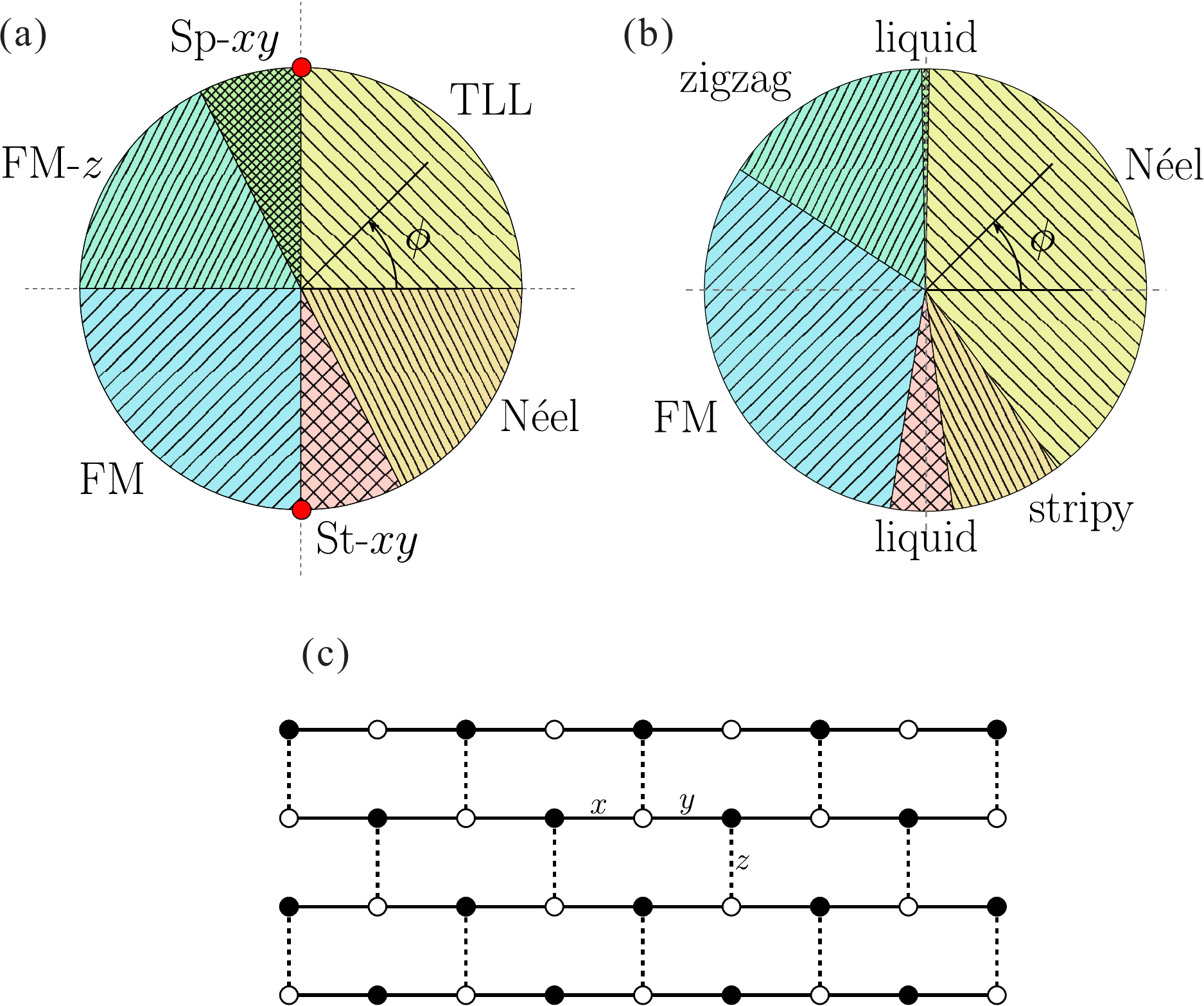}
\caption{
(a) Ground-state phase diagram of the 1D KH model. (b) Ground-state phase diagram of the honeycomb-lattice KH model, where the same notations as in Eq.~(\ref{ham}) are used for $K$ and $J$. (c) Brick-wall lattice created by coupling the 1D KH chains by the $z$-bond. This is topologically equivalent to the honeycomb lattice.
}
\label{pd}
\end{figure}

In Fig.~\ref{pd}(a) we summarize the ground-state phase diagram as a function of $\phi$ in a pie chart form. Notably, each phase has another exactly-symmetrically-placed phase in the pie chart, i.e, TLL and FM-$xy$, N\'eel-$z$ and FM-$z$, spiral-$xy$ and staggered-$xy$. The paired phases have the common features. The TLL and FM-$xy$ states, where the exchange components are either all AFM or all FM, are basically understood within the framework of the isotropic Heisenberg chain; the FM-$xy$ state is long-range ordered and the TLL state is critical. The N\'eel-$z$ and FM-$z$ states are described by the easy-axis XXZ Heisenberg chain. The dominant features are determined by the Ising term. Depending on the sign of the Ising term, the spins are aligned ferromagnetically or antiferromagnetically along the $z$ axis. The spiral-$xy$ and staggered-$xy$ states can be basically interpreted as the easy-plane XXZ Heisenberg chain affected by the double-spin-flip term. The system has a four-site periodicity, and exhibits a critical behavior and a long-range ordering for the former and latter states, respectively, in analogy to the relation between the TLL and FM-$xy$ states.

It is now possible to establish a connection between our ground-state phase diagram and that of the two-dimensional KH model on a honeycomb lattice [Fig.~\ref{pd}(b), normalized to the current notation from \cite{Chaloupka10}]. To consider the phase-to-phase correspondence, it is helpful to regard the honeycomb lattice as coupled KH chains, as shown in Fig.~\ref{pd}(c). Here, the interchain coupling is equivalent to the $z$-bond in the honeycomb-lattice  KH model, written as
\begin{equation}
{\cal H}_{z-{\rm bond}}=\frac{J}{2}\sum_{k,l} (S^+_k S^-_l+S^-_k S^+_l)+(J+K)\sum_{k,l} S^z_k S^z_l,
\label{zbond}
\end{equation}
where $k$, $l$ are summed over all connected sites between the KH chains. 

Hereafter, we report how the introduction of the $z$-bonds coupling affects the (quasi-)ordered states present in the one-dimensional case, mapping them to those of the honeycomb system:

(i) At $0\le\phi<\frac{\pi}{2}$ the interchain coupling is AFM. Our TLL state has no long-range order but strong AFM fluctuations. Therefore, a N\'eel order is {intuitevely} expected once the chains are antiferromagnetically coupled. {Hence,} our TLL phase corresponds to the N\'eel phase in the honeycomb case. 

(ii) At $0.65\pi\lesssim\phi<\pi$ our system is in the FM-$z$ state. {While considering this interval, we need to examine two different cases as the sign of the interaction on the $z$-bonds changes from AFM to FM.} We obtain a zigzag state in the honeycomb case when {the FM-$z$ chain} state is coupled by AFM interchain couplings; whereas, a FM state in the case of FM interchain couplings. The {change in sign} of the interchain coupling is assumed {to happen} at $\phi=\frac{3\pi}{4}$, {where} $J+K=0$. This $\phi$ value is reasonably close to the transition point $\phi\approx0.81\pi$ between the zigzag and FM states in the honeycomb case. 
{Thus}, our FM-$z$ phase is distributed to either zigzag or FM phases of the honeycomb KH model depending on the AFM or FM nature on the $z$-bond. 

(iii) At $1.65\pi\lesssim\phi<2\pi$ our system is in the N\'eel-$z$ state. In the same way as above, {we need to consider the cases of AFM and FM $z$-bonds.} We easily find that our N\'eel-$z$ state is base for the N\'eel and stripy states of the honeycomb KH model as also discussed in previous work \cite{Sela14}. Namely, our N\'eel-$z$ phase is distributed to either N\'eel  phase {in the presence of AFM interchains coupling}
 or the stripy phase {in the presence of FM interchains coupling} of the honeycomb KH model. The value of $\phi=\frac{7\pi}{4}$ giving $J+K=0$ is again near the transition point $\phi\approx1.65\pi$ between the N\'eel and stripy states in the honeycomb case. 

Therefore, it is possible to understand all the long-range ordered phases of the honeycomb-lattice KH model in the framework of the coupled KH chains. {In brief, an extracted zigzag lattice line from any ordered state of the honeycomb-lattice KH model corresponds to one of the phases in the 1D KH chain; and the remaining degrees of freedom derive from the fact that the zigzag lines are coupled either ferromagnetically or antiferromagnetically by ${\cal H}_{z-{\rm bond}}$.}

(iv) Two $xy$ phases in our phase diagram are left. Since these states are stabilized by the dominant {$xy$-Kitaev term}, they are connected to the Kitaev spin liquid states in the honeycomb case. {In hindsight}, this means that our staggered-$xy$ ordered state can collapse {to the spin liquid state} due to strong FM fluctuations along the $z$-direction.

\subsection*{Low-lying excitations} 

\begin{figure}[!t]
\centering
\includegraphics[width=0.50\linewidth]{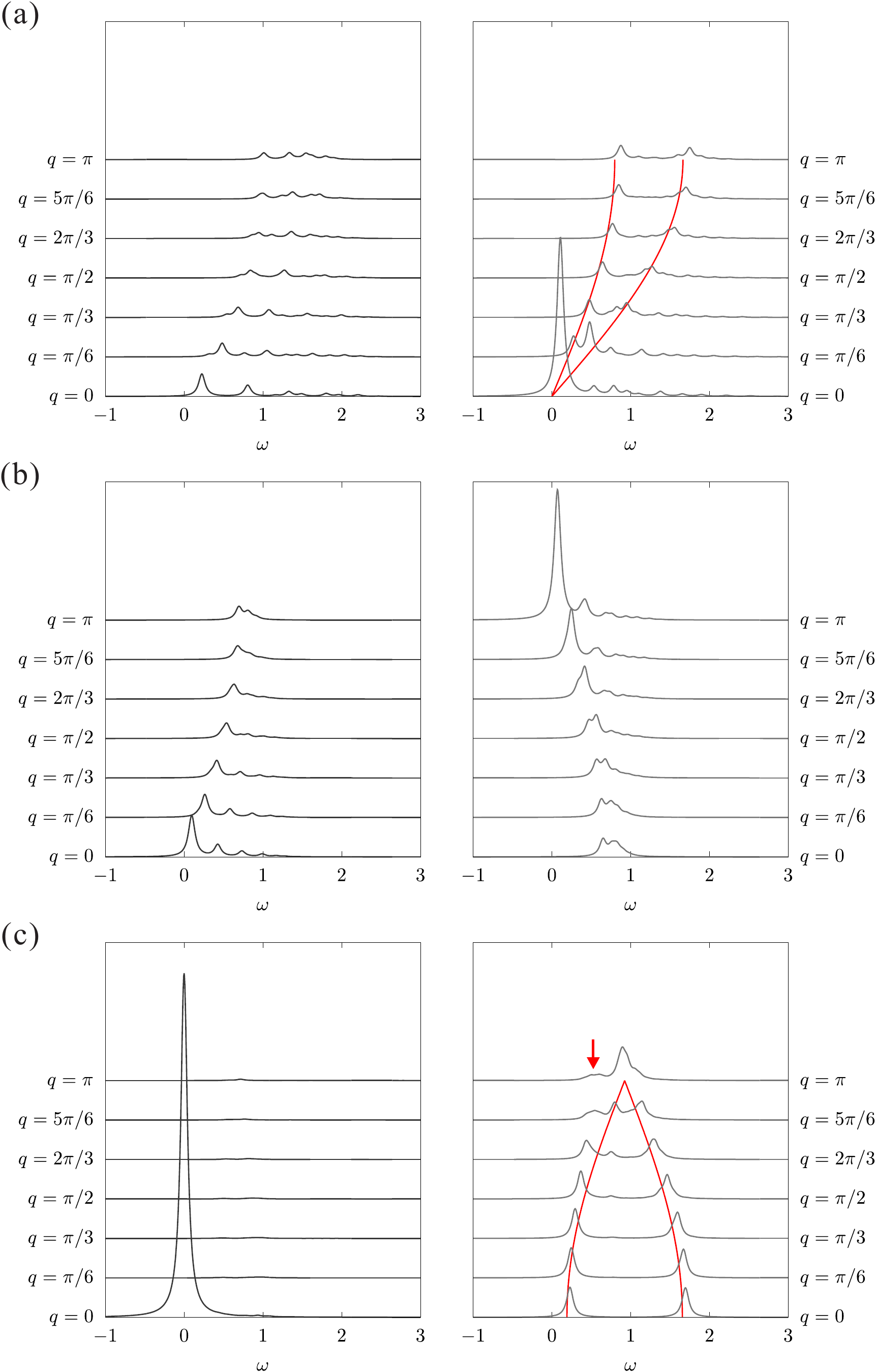}
\caption{
Dynamical structure factors calculated with a periodic 24-site chain for (a) TLL [$\phi=\frac{\pi}{3}$], (b) spiral-$xy$ [$\phi=\frac{5\pi}{8}]$, and (c) FM-$z$ [$\phi=\frac{7\pi}{8}$] states. The left and right panels show $S^z(q,\omega)$ and $S^-(q,\omega)$, respectively. The red lines are analytical dispersions: Eq.~(\ref{omega_TL}) for $\phi=\frac{\pi}{3}$ and Eq.~(\ref{omega_FMz}) for $\phi=\frac{7\pi}{8}$. The arrow indicates an appearance of the spiral-$xy$ fluctuation in the FM-$z$ state.
}
\label{Sqw1}
\end{figure}

\begin{figure}[!t]
\centering
\includegraphics[width=0.50\linewidth]{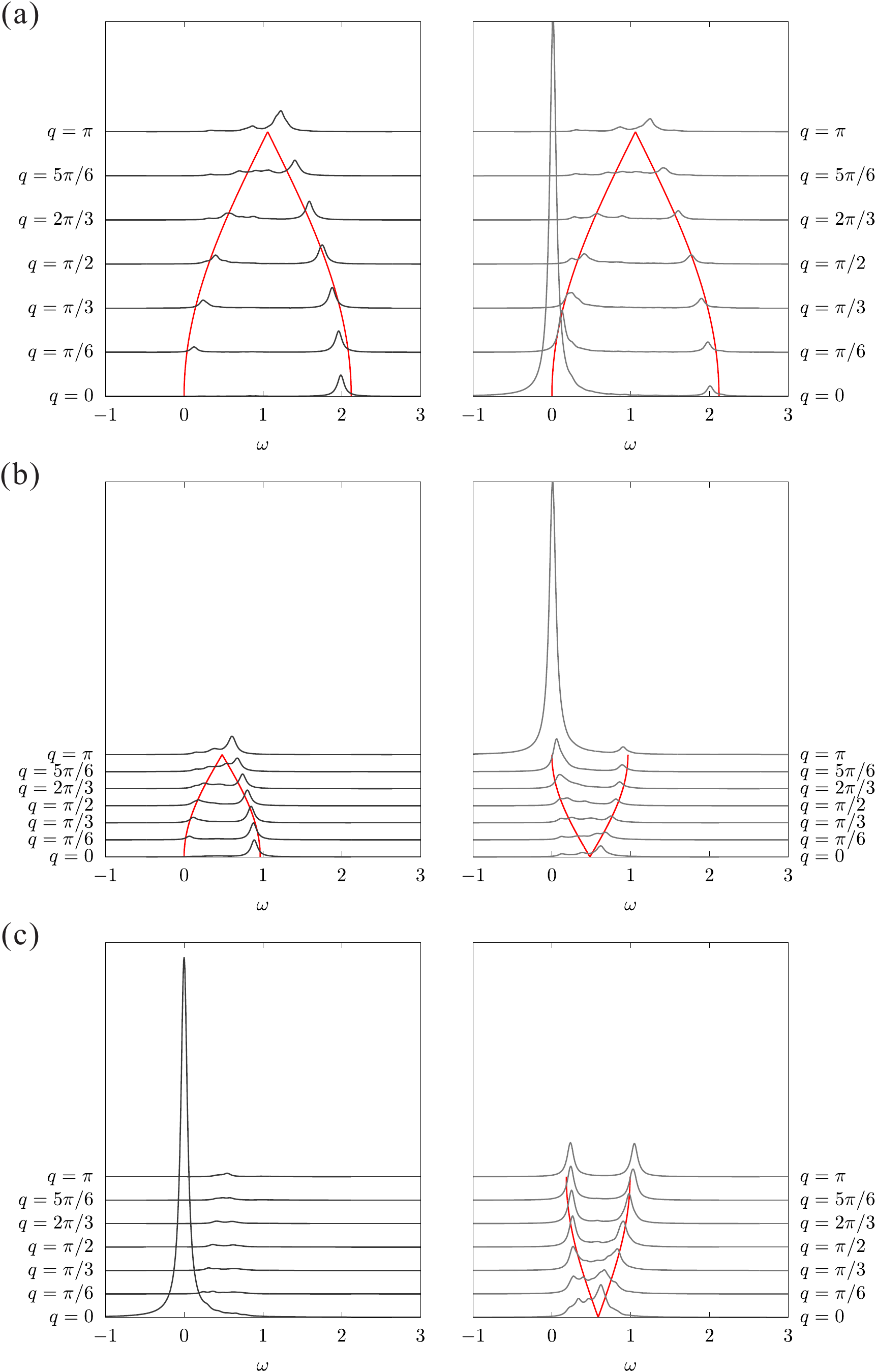}
\caption{
Dynamical structure factors calculated with a periodic 24-site chain for (a) FM-$xy$ [$\phi=\frac{5\pi}{4}$], (b) staggered-$xy$ [$\phi=\frac{19\pi}{12}$], and (c) N\'eel-$z$ [$\phi=\frac{17\pi}{10}$] states. The left and right panels show $S^z(q,\omega)$ and $S^-(q,\omega)$, respectively. The red lines are analytical dispersions: Eq.~(\ref{omega_FMxy}) for $\phi=\frac{5\pi}{4}$, Eq.~(\ref{omega_XY1}) in $S^z(q,\omega)$ and Eq.~(\ref{omega_XY2}) in $S^-(q,\omega)$ for $\phi=\frac{19\pi}{12}$, and Eq.~(\ref{omega_Neel1}) for $\phi=\frac{17\pi}{10}$.
}
\label{Sqw2}
\end{figure}

To examine the low-energy excitations for each phase, we calculated the dynamical spin structure factors $S^{z,\pm}(q,\omega)$, defined in Eq.~(\ref{spec}). Under periodic boundary conditions the momentum is taken as $q=\frac{4\pi}{L}n$ ($n=0,\pm1,\dots, \pm\frac{L}{4}$) since the unit cell contains two lattice sites and the number of unit cells is $\frac{L}{2}$ in a system with $L$ sites. We study chains with $L=24$ using the Lanczos ED method and the results are shown only for $q\ge0$ as $S^{z,\pm}(q,\omega)=S^{z,\pm}(-q,\omega)$.

\subsubsection*{Tomonaga-Luttinger-liquid phase ($0 \le \phi <\pi/2$)} 

As confirmed above, the low-energy physics at $0 \le \phi <\frac{\pi}{2}$ is described by a gapless TL model. The low-lying excitations are expected to be basically equivalent to those of the easy-plane XXZ chain because $(2J+K)/2>J>0$ [see Eq.~(\ref{ham3})]. If we ignore the double-spin-flip term ${\cal H}_{\rm dsf}$, the main dispersion (lower bound of the spectrum) is given by
\begin{equation}
\omega(q) = \frac{\pi(2J+K)}{4}\frac{\sin\mu}{\mu}\sin\frac{q}{2}
\label{xxz_omega}
\end{equation}
with $\cos\mu=\left(1+\frac{K}{2J}\right)^{-1}$. This was obtained by using the transfer matrix of the six-vertex model~\cite{Johnson73,Schneider82}. Nevertheless, the effect of ${\cal H}_{\rm dsf}$ is unknown. Therefore, to investigate the effect of ${\cal H}_{\rm dsf}$ we employ a standard SWT for the bipartite system. Using the Holstein-Primakovff representation the spin operators of ${\cal H}_{\rm dsf}$ for the black labeling sites in Fig.\ref{fig_lattice}(d) are replaced as $S^+_i=\sqrt{2S-a^\dagger_ia_i}a_i$ and $S^-_i=a_i^\dagger\sqrt{2S-a^\dagger_ia_i}$; similarly, for the white labeling sites $S^+_i=b_i^\dagger\sqrt{2S-b^\dagger_ib_i}$ and $S^-_i=\sqrt{2S-b^\dagger_ib_i}b_i$, where $a_i$ and $b_i$ are canonical boson annihilation operators at site $i$. Then, applying the Fourier transform we obtain
\begin{equation}
{\cal H}_{\rm dsf}=i\frac{K}{2}\sum_q\sin\frac{q}{2}(a_\frac{q}{2}^\dagger b_\frac{q}{2}-a_\frac{q}{2}b_\frac{q}{2}^\dagger).
\label{dsf_SWT}
\end{equation}
This off-diagonal term gives a splitting in the main dispersion (\ref{xxz_omega}). Then, we can speculate the total dispersions as
\begin{equation}
\omega(q) = \left(\frac{\pi(2J+K)}{4}\frac{\sin\mu}{\mu}\pm\frac{K}{2}\right)\sin\frac{q}{2}.
\label{omega_TL}
\end{equation}
Fig.~{\ref{Sqw1}}(a) shows the dynamical structure factors $S^z(q,\omega)$ and $S^-(q,\omega)$ for $\phi=\frac{\pi}{3}$. The largest peak appears in $S^-(q=0,\omega=0)$ reflecting the AFM fluctuations. The intensities in $S^z(q,\omega)$ are weaker than those in $S^-(q,\omega)$ due to the easy-plane $xy$ anisotropy. The split dispersions are well described by Eq.~(\ref{omega_TL}). The splitting is largest in the vicinity of the Kitaev point $\phi=\frac{\pi}{2}-$ ($J\to0+$, $K\to1$):
\begin{equation}
\omega(q) = \left(\frac{K}{2}\pm\frac{K}{2}\right)\sin\frac{q}{2}.
\label{omega_Kitaev}
\end{equation}
 The main dispersion (lower bound of the spectrum) of the isotropic Heisenberg model at $\phi=0$ ($K=0$) is also reproduced by  Eq.~(\ref{omega_TL}).

\subsubsection*{Spiral-$xy$ phase ($\frac{\pi}{2}<\phi\lesssim0.65\pi$)} 

Across the Kitaev point from the TL to spiral-$xy$ phases the momentum of the Fermi point is shifted from $q=\pi$ to $q=0$. Thus, in the vicinity of the Kitaev point $\phi=\frac{\pi}{2}+$ ($J\to0-$, $K\to1$), the momentum of the excitation dispersion is shifted from Eq.~(\ref{omega_Kitaev}) by $\pi$:
\begin{equation}
\omega(q) = \left(\frac{K}{2}\pm\frac{K}{2}\right)\left(1-\sin\frac{q}{2}\right)
\label{omega_Kitaev2}
\end{equation}
for $S^-(q,\omega)$. In contrast, $S^z(q,\omega)$ exhibits a $q$-independent continuum between $\omega=0$ and $\omega=K$ since the $z$-component of exchange interaction is much smaller than the other interactions. With increasing $\phi$ from $\frac{\pi}{2}$, the FM Ising interaction and double-spin-flip fluctuations increase and the exchange interaction decreases becoming zero at the critical boundary to the FM-$z$ phase. Fig.~\ref{Sqw1}(b) shows the dynamical structure factors $S^z(q,\omega)$ and $S^-(q,\omega)$ for $\phi=\frac{5}{8}\pi$. The main dispersion in $S^-(q,\omega)$ is scaled as $\omega(q) \propto 1-\sin\frac{q}{2}$. Although the largest peak at $(q,\omega)=(\pi,0)$ indicates a four-site periodicity in the spiral-$xy$ state, it diminishes with increasing system size because of no long-range ordering. While in $S^z(q,\omega)$, a dispersion $\omega(q) \propto \sin\frac{q}{2}$ appears and a peak toward the FM-$z$ state develops at $(q,\omega)\approx(0,0)$. Both the dispersion widths are roughly scaled by $|K|$.

\subsubsection*{Ferromagnetic-$z$ phase ($0.65\pi\lesssim\phi<\pi$)}

Fig.~\ref{Sqw1}(c) shows the dynamical structure factors $S^z(q,\omega)$ and $S^-(q,\omega)$ for $\phi=\frac{7}{8}\pi$. Near $\phi=\pi$ ($K\approx0$) in the FM-$z$ phase, the system is effectively described by a FM XXZ Heisenberg chain with easy-axis anisotropy $J<\frac{2J+K}{2}<0$ and the ground state is approximately expressed by Eq.~(\ref{fm2}). An applied operator $S^z_q$ does not change the ground state $| \psi_0 \rangle$ in Eq.~(\ref{spec}) and the final state $|\psi_\nu \rangle$ has only a zero energy excitation from $| \psi_0 \rangle$. Therefore, $S^z(q,\omega)$ has a very sharp peak at $(q,\omega)=(0,0)$ and almost no spectral weight appears at the other momenta. This peak keeps its weight constant with increasing system length, indicating the long-range FM-$z$ ordering. On the other hand, in Eq.~(\ref{spec}) the operator $S^-_q$ dopes one magnon into the FM-$z$ alignment so that the SWT is expected to give a good approximation for the excitation dispersion of $S^-(q,\omega)$:
\begin{equation}
\omega(q) = -J \pm \frac{2J+K}{2}\cos\frac{q}{2}.
\label{omega_FMz}
\end{equation}
We can confirm that a gap $\Delta=\frac{K}{2}$ opens at $q=0$ reflecting the easy-axis anisotropy. With approaching the neighboring spiral-$xy$ phase, the double-spin-flip fluctuations grow gradually in influence; accordingly, the $q=0$ peak in $S^z(q,\omega)$ shrinks and a peak develops at $q=\pi$, $\omega\approx0$ in $S^-(q,\omega)$.

\subsubsection*{Ferromagnetic-$xy$ phase ($\pi<\phi<\frac{3\pi}{2}$)}

Fig.~\ref{Sqw2}(a) shows the dynamical structure factors $S^z(q,\omega)$ and $S^-(q,\omega)$ for $\phi=\frac{5}{4}\pi$. Near $\phi=\pi$ ($K\approx0$) in the FM-$xy$ phase, the system is effectively described by a FM XXZ Heisenberg chain with easy-plane anisotropy $\frac{2J+K}{2}<J<0$ and the ground state is approximately expressed by Eq.~(\ref{fm1}). Thus, the excitation spectrum is expected to be gapless. Since the total $S^z$ of the ground state is zero, unlike the case of FM-$z$ state both $S^z(q,\omega)$ and $S^-(q,\omega)$ have the same excitation dispersion as
\begin{equation}
\omega(q) = -\frac{2J+K}{2}\left(1\pm\cos\frac{q}{2}\right).
\label{omega_FMxy}
\end{equation}
Note that a sharp peak at $(q,\omega)=(0,0)$ in $S^-(q,\omega)$ keeps its weight constant with increasing system length, indicating the  FM-$xy$ long-range ordering. On the other hand, in the vicinity of the Kitaev point $\phi\approx\frac{3\pi}{2}-$, the excitation dispersion of $S^-(q,\omega)$ is well described by Eq.~(\ref{omega_Kitaev}).

\subsubsection*{Staggered-$xy$ ordered phase ($\frac{3\pi}{2}<\phi\lesssim1.65\pi$)} 

In the vicinity of the Kitaev point $\phi=\frac{3}{2}\pi+$, the excitation spectra are the same as in the other Kitaev point $\phi=\frac{1}{2}\pi+$, namely, Eq.~(\ref{omega_Kitaev2}) for $S^-(q,\omega)$ and a $q$-independent continuum for $S^z(q,\omega)$. Whereas in the vicinity of the N\'eel-$z$ ordered phase $\phi=\tan^{-1}(-2)-$, the ground state is expressed by Eq.~(\ref{XY1}). Thus, the dispersions are described by a single magnon excitation:
\begin{equation}
\omega(q) = -\frac{K}{2} \left(1\pm \cos\frac{q}{2}\right)
\label{omega_XY1}
\end{equation}
for $S^z(q,\omega)$, and
\begin{equation}
\omega(q) = -\frac{K}{2} \left(1\pm \sin\frac{q}{2}\right)
\label{omega_XY2}
\end{equation}
for $S^-(q,\omega)$. Fig.~\ref{Sqw2}(b) shows the dynamical structure factors $S^z(q,\omega)$ and $S^-(q,\omega)$ for $\phi=\frac{19\pi}{12}$. The excitation dispersions are well reproduced by Eqs.~(\ref{omega_XY1}) and (\ref{omega_XY2}). We also find a sharp peak at $(q,\omega)=(\pi,0)$ in $S_-(q,\omega)$, which indicates the staggered-$xy$ long-range ordering. This peak keeps its weight constant with increasing the system length, in contrast to the similar peak for the spiral-$xy$ state.

\subsubsection*{N\'eel-$z$ ordered phase ($1.65\pi\lesssim\phi<2\pi$)} 

In the vicinity of the staggered-$xy$ ordered phase $\phi=\tan^{-1}(-2)+$, the ground state is expressed by Eq.~(\ref{Neel1}). Accordingly, $S^z(q,\omega)$ has only a delta peak at $(q,\omega)=(0,0)$. Whereas, $S^-(q,\omega)$ is exactly explained by a single magnon dispersion. It is obtained by the SWT as
\begin{equation}
\omega(q) = J \pm \frac{K}{2}\sin\frac{q}{2}.
\label{omega_Neel1}
\end{equation}
Although this is equivalent to Eq.~(\ref{omega_XY1}), the spectral weight is uniform for all $q$ values. Since the transition at $\phi=\tan^{-1}(-2)$ is  first ordered, there is no peak indicating a connection to the staggered-$xy$ ordered phase. On the other hand, near $\phi=2\pi$, the system can be basically regarded as an easy-axis AFM XXZ Heisenberg chain so that the excitation dispersion is described by Eq.~(\ref{omega_TL}) for both $S^z(q,\omega)$ and $S^-(q,\omega)$. Fig.~\ref{Sqw2}(c) shows the dynamical structure factors $S^z(q,\omega)$ and $S^-(q,\omega)$ for the intermediate region $\phi=\frac{17}{10}\pi$. The main dispersion of $S^-(q,\omega)$ is basically described by Eq.~(\ref{omega_Neel1}) but some features from Eq.~(\ref{omega_TL}) seem to be somewhat mixed.

\section*{Discussion}

We have established the presence of a variety of phases in the 1D KH system. Especially, it is surprising that most of the $\phi$ ranges are covered by long-range ordered phases despite considering a pure 1D system. In this context, we now consider the K-intercalated $\alpha$-RuCl$_3$, namely, K$_{0.5}$RuCl$_3$. One should be aware of the fact that several different parameter sets have been suggested for undoped $\alpha$-RuCl$_3$: (i) $K=-5.6$, $J=1.2$ ($\phi\approx1.57\pi$)~\cite{Ravi16}, (ii) $K=7.0$ ,$J=4.6$ ($\phi\approx0.68\pi$)~\cite{Banerjee16}, (iii) $K=8.1$, $J=2.9$ ($\phi\approx0.61\pi$)~\cite{Banerjee16}, (iv) $K=-6.8$, $J=0$ ($\phi=1.5\pi$)~\cite{Ran17} in unit of meV. If we assume that the charge ordering pattern in K$_{0.5}$RuCl$_3$ is that illustrated in Fig.~\ref{fig_lattice}(c), the parameter sets (i)-(iv) correspond to the staggered-$xy$, FM-$z$, spiral-$xy$, and FM Kitaev point, respectively. In practice, the charge ordering could cause a significant change of the parameter since they are very sensitive to the Ru-Cl-Ru bond angle~\cite{Ravi16}. In other words, once the magnetic properties of K$_{0.5}$RuCl$_3$ are observed, we may easily speculate the possible parameter set of K$_{0.5}$RuCl$_3$ and even the charge ordering pattern by comparing then to our rich phase diagram. To gain deeper insights, theoretical and experimental studies under magnetic fields are also required.

It is relevant to seek other possible realizations of 1D KH system. Even if the Kitaev interaction in 1D systems is present, it is considerede to be very small. However, as shown above, even a tiny Kitaev interaction can stabilize the ordered state. It might also be intriguing to reconsider quasi-1D materials having two sublattices like Ni$_2$(EDTA)(H$_2$O)$_4$, [Ni(f-rac-L)(CN)$_2$], LiCuSbO$_4$, and  Rb$_2$Cu$_2$Mo$_3$O$_{12}$ from the point of view of the 1D Kitaev system.

The $\phi$-dependent phase diagram of our model is similar to that for the honeycomb-lattice KH model. Remarkably, all the magnetically ordered states of the honeycomb-lattice KH model can be interpreted in terms of the coupled 1D KH chains. In other words, the key elements to derive the magnetic ordering in the 2D honeycomb-lattice KH model are already contained in the 1D KH chain. This will give us a deeper insight in the understanding of transition from an ordered state to a disordered state such as the Kitaev QSL.

\section*{Methods}

In this section, we present a detailed derivation of (\ref{bwspin1}), (\ref{bwspin2}), and the related additional information. We used a Jordan-Wigner transformation
\begin{align}
	S_i^z&=2c_i^\dagger c_i -1 \\
	S_i^+&=\prod_{k=1}^{i-1}(-S^z_k)c_i^\dagger \\
	S_i^-&=\prod_{k=1}^{i-1}(-S^z_k)c_i 
\end{align}
to rewrite the spin operators in Eq.~(\ref{eq:KHam}) by fermion operators, and we obtained Eq.~(\ref{eq:FermHam}). 

By applying the inverse of the Jordan-Wigner transformation
\begin{align}
	c_{w,i}^\dagger &= \prod_{k=1}^{i-1}\prod_{k'=b,w}(-S^z_{k,k'})(-S^z_{b,i}) S_{w,i}^+ \\
	\label{cwi}
	c_{w,i} &= \prod_{k=1 }^{i-1}\prod_{k'=b, w}(-S^z_{k,k'})(-S^z_{b,i}) S_{w,i}^- \\
	c_{b,i}^\dagger &= \prod_{k=1 }^{i-1}\prod_{k'=b, w}(-S^z_{k,k'}) S_{b,i}^+ \\
	c_{b,i} &= \prod_{k=1 }^{i-1}\prod_{k'=b, w}(-S^z_{k,k'}) S_{b,i}^-,
	\label{cbi}
\end{align}
the fermion operator $\widetilde{d}_i^{\, \dagger}$ was traced back to the initial spin operators, and we obtained Eq.~(\ref{tddag2}) from Eq.~(\ref{tddag1}).

We can also derive the same results as Eqs.~(\ref{bwspin1}) and (\ref{bwspin2}) by taking a different combination of the Majoranas from Eqs.~(\ref{recomb1}) and (\ref{recomb2}):
\begin{align}
	B_i^\dagger&=\frac{b_{1,i}-ib_{1,i+1}}{2} 	\label{eq:Bferm} \\
	W_i^\dagger& = \frac{w_{2,i}-iw_{2,i+1}}{2} \label{eq:Wferm}
\end{align}
From Eqs.~\eqref{eq:Ferm1} and \eqref{eq:Ferm2}, we have
\begin{align}
	B_i^\dagger&=\frac{1}{2}(c_{b,i}^\dagger + c_{b,i} -i c_{b,i+1}^\dagger -ic_{b,i+1}) \\
	W_i^\dagger&=\frac{1}{2}(c_{w,i+1}^\dagger - c_{w,i+1} + ic_{w,i}^\dagger - i c_{w,i})
\end{align}
Using (\ref{cwi})-(\ref{cbi}), we obtain
\begin{align}
	B_i^\dagger&= \prod_{k=1}^{i-1} \prod_{k'=b,w} (-S_{k',k})\left[S_{b,i}^x - i (-S_{k',i}^z)S_{b,i+1}^x \right]\\
	B_i&= \prod_{k'=b,w} (-S_{k',k})(S_{b,i}^x - i (-S_{k',i}^z)S_{b,i+1}^x) \label{eq:Bspin}
\end{align}
Then, we obtain
\begin{equation}
	b_{1,i}=B_i^\dagger+B_i=2\prod_{k=1}^{i-1}(-S_{k,k'}^z)S_{b,i}^x. \label{eq:b1i}
\end{equation}
In a similar way, we find
\begin{equation}
	w_{2,i}=W_i^\dagger+W_i=2\prod_{k=1}^{i-1} \prod_{k'=w,b} (-S_{k',k}^z)(-S_{b,i}^z)(-S_{w,i}^y).\label{eq:w2i}
\end{equation}
Actually, Eqs.~(\ref{eq:b1i}) and (\ref{eq:w2i}) are identical to Eqs.~(\ref{bwspin1}) and (\ref{bwspin2}).

%\bibliographystyle{naturemag-doi}

%\noindent LaTeX formats citations and references automatically using the bibliography records in your .bib file, which you can edit via the project menu. Use the cite command for an inline citation, e.g.  \cite{Figueredo:2009dg}.

\section*{Acknowledgements}

We thank U. Nitzsche for technical assistance. C.E.A. thanks A. Lau for useful discussion. This work is supported by SFB 1143 of the Deutsche Forschungsgemeinschaft.

%\section*{Author contributions statement}

%C.E.A performed the numerical calculations and analysed the results with assistance from S.N. All authors reviewed the manuscript. 

%\section*{Additional information}

%To include, in this order: \textbf{Accession codes} (where applicable);

%\textbf{Competing financial interests} The authors declare no competing financial interests. 

%The corresponding author is responsible for submitting a \href{http://www.nature.com/srep/policies/index.html#competing}{competing financial interests statement} on behalf of all authors of the paper. This statement must be included in the submitted article file.

%Figures and tables can be referenced in LaTeX using the ref command, e.g. Figure \ref{fig:stream} and Table \ref{tab:example}.

\end{document}